\documentstyle[prb,aps,psfig]{revtex}

\begin{document}

\title{Mean-field approach to ferromagnetism in (III,Mn)V diluted
magnetic semiconductors at low carrier densities}
 
\author{Mona Berciu and R. N. Bhatt}

\address{Department of Electrical Engineering, Princeton University,
Princeton, New Jersey 08544} \date{\today}

\twocolumn[\hsize\textwidth\columnwidth\hsize\csname@twocolumnfalse\endcsname

\maketitle
\begin{abstract}
We present a detailed study, within the mean-field approximation, of
an impurity band model for III-V diluted magnetic semiconductors. Such
a model should be relevant at low carrier densities, below and near
the metal-insulator transition. Positional disorder of the magnetic
impurities inside the host semiconductor is shown to have observable
consequences for the shape of the magnetization curve.  Below the
critical temperature the magnetization is spatially inhomogeneous,
leading to very unusual temperature dependence of the average
magnetization as well as specific heat. Disorder is also found to
enhance the ferromagnetic transition temperature. Unusual spin and
charge transport is implied.
\end{abstract}
 ]

\section{Introduction}

Diluted magnetic semiconductors (DMS) are semiconductors of the
general type A$_{1-x}$M$_x$B, where AB is either a II-VI or a III-V
semiconductor and M a magnetic element, most commonly Mn. Substitution
of a small fraction $x$ of the element A by Mn impurities (and in the
case of II-VI semiconductors an additional charge dopant, such as P on
the B site) leads to the emergence of a semiconductor with
ferromagnetic properties.\cite{Ohnorev} This is due to the
interactions of the $S={5 \over 2}$ Mn spins (coming from the
half-filled 3d shell of the Mn) with the spins of the charge carriers
introduced by the Mn dopants, or, in the case of II-VI semiconductors,
by the additional dopant.  This opens up the possibility of
manipulating (through doping) not only the charge, but also the spin
properties of the semiconductor.  The ability to control the
properties of a system by acting on the spin of its charge carriers is
the subject of the new field of ``spintronics'', and is believed to
hold the promise to developing devices which combine storage
functionalities (such as memory devices) together with information
processing functionalities.

The recent demonstration of Curie temperatures of the order of 100 K
in Ga$_{1-x}$Mn$_x$As samples with $x \approx 0.05$, grown at
low-temperatures using molecular-beam epitaxy (MBE) techniques,
\cite{Ohno,Haya,Esch} has heightened the interest in understanding the
physics of these alloys. By now, it is well established that the main
magnetic interaction is an antiferromagnetic (AFM) exchange between
the Mn spins and the charge carrier spins. As a result, an effective
ferromagnetic (FM) interaction arises between the Mn spins through
carrier-induced ferromagnetism. Several theories, most notably
Ruderman-Kittel-Kasuya-Yosida (RKKY),\cite{RKKY} have been used to
explain, within a mean-field approximation, this phenomenon. More
recently, dynamic correlations as well as arbitrary itinerant-carrier
spin polarizations have been included.\cite{MacD} However, all these
models assume that the charge carriers (the holes) occupy a Fermi sea
in the valence band. This implicitly assumes a spatially homogeneous
distribution of the holes throughout the entire system, and therefore
completely neglects the role of the Coulomb attraction between the
charge carriers and the Mn ions, as well as the role of disorder which
is always present in such alloy systems.

It is well known \cite{BG} that Mn impurities in a III-V semiconductor
create a trapping potential for holes. The associated acceptor levels
are about 100 meV above the top of the valence band, for GaAs, and
interactions will lead to their splitting into an impurity
band.\cite{SE,BhattRice81} At $T=0$, the holes occupy the states of
lowest energy and therefore they first occupy states in the impurity
band. Only if the Fermi energy (or thermal energy $k_BT$) is large
enough are states in the valence band occupied as well. It has been
found experimentally that these alloys are heavily compensated,
leading to rather small hole concentrations, of the order of 10\% of
the Mn concentration.\cite{Besch} The corresponding small Fermi energy
implies a long screening length for the Coulomb interactions, and
opens up the possibility that holes are actually moving (through
hopping processes) in the impurity band formed of states localized
about the Mn impurities. Strong experimental evidence for this
scenario comes from electrical conductivity measurements, which at low
temperatures reveal Mott variable-range hopping
behavior. \cite{Esch,Katsu} As a result, randomness in the position of
the Mn spins (and associated random potential) could be expected to
play an important role.

In this paper we investigate, at the mean-field level, a model in
which holes move in a band formed of impurity states, neglecting the
existence of the valence band states.  Some results of this model have
already been reported in Ref. 12.  While a complete model should
include both impurity and valence band states, we believe that our
model is a good first approximation, especially at low temperatures,
since we find typical Fermi energies of a few tens of meV, smaller
than the average $\sim 100 $meV between the impurity band and the
valence band.

The paper is organized as follows. In Section II we describe the model
Hamiltonian, the self-consistent mean-field approximation as well as
the specific parameters used in the numerical calculations. It is
important to emphasize that although we chose numbers from the
literature specific to the GaMnAs problem for illustration purposes,
similar arguments and physics would hold for other III-V compounds. In
Section III we present our results. We first analyze the hypothetical
case in which all Mn impurities are ordered in a simple cubic
lattice. This allows us to understand the unusual shape of the
magnetization curves we obtain. More importantly, it allows us to
clearly identify the non-trivial effects of randomness (disorder) in
the Mn positions. These are analyzed in some detail and a clear
physical picture of their importance to the problem emerges. We also
analyze the metal-insulator transition, the effects of on-site random
disorder, as well as other possible interactions. The sensitivity of
our results to variations in the impurity band parameters is also
discussed.  Finally, Section IV contains a summary and conclusions.

\section{Model}

Based on experimental work in the literature,\cite{Ohnorev} we assume
 that Mn only substitutes for the group III element of the III-V
 semiconductor. The III-V semiconductor is assumed to have a
 zinc-blende structure. Let $\vec{R}_i, i=1,...,N_d$ be the random
 positions of the $N_d$ Mn dopants in the face centered cubic (FCC)
 sublattice of the group III element.  Each Mn impurity is associated
 with a spin-${5 \over 2}$ from its half-filled $3d$ shell. Since Mn
 has nominal valence II, when it substitutes for the valence-III
 element it will act as an acceptor. As a result, it can trap a hole
 shallow level, characterized by a hydrogenic Bohr radius $a_B$.  Let
 $N_h$ be the total number of holes trapped about various Mn sites,
 and $p=N_h /N_d$ the relative hole concentration. A hole can hop from
 one Mn impurity to another one, while its spin is
 antiferromagnetically coupled to the Mn spins in its vicinity.

For simplicity, although the charge carriers of this system are holes,
in the following we use an ``electron''-formalism to analyze them. In
other words, we in fact analyze an equivalent system doped with
hypothetical donors, with impurity levels below a conduction-like
band, instead of being above a valence-like band. The main difference
between the two models is (i) the hole spin is ${3\over 2}$ and (ii)
the envelope wave function of hole has cubic symmetry, rather than
spherical symmetry of a donor. This leads to some quantitative
differences (as has been shown by MacDonald and coworkers for free
holes in Ref. 13) but the essential aspects of the problem, namely (a)
disorder and (b) the effect of impurity potentials, which we
concentrate on here, remain substantially unaffected.

The Hamiltonian we study is:
$${\cal H}=\sum_{i,j}^{}t_{ij}c^{\dagger}_{i\sigma} c_{j\sigma}
+\sum_{i}^{}u(i) c^{\dagger}_{i\sigma} c_{i\sigma}
$$ $$ +\sum_{i,j}^{} J_{ij} \vec{S}(i) \left(c^{\dagger}_{j\alpha} {1
\over 2} \vec{\sigma}_{\alpha\beta} c_{j\beta}\right)
$$
\begin{equation}
\label{1}
- g \mu_BH\sum_{i}^{}{\sigma \over 2} c^{\dagger}_{i\sigma}
  c_{i\sigma} - \tilde{g} \mu_B H \sum_{i}^{}S^z(i).
\end{equation}
Here, $c^{\dagger}_{i\sigma}$ is the creation operator of a charge
carrier with spin $\sigma$ in the bound orbital associated with the
$i$th Mn impurity.  The first term in Eq. (\ref{1}) describes charge
carriers hopping between various Mn sites, with the hopping matrix
$t_{ij}$ dependent on the hopping distance
$r=|\vec{R}_i-\vec{R}_j|$. We take an exponential form like for
hydrogenic orbitals $ t(r) = 2\left( 1 + { r / a_B}\right)
\exp{\left(-{r/a_B}\right)} \mbox{ Ry} $. \cite{Bhatt1} We present
detailed results for this case; however we discuss in subsection F the
changes that result with other models of the hopping integral. From
these different cases we discern the universal aspects of the impurity
models and also the requisite conditions for the existence of the
experimentally observed ferromagnetism.

The Rydberg (Ry) is defined by the binding energy of the charge
carrier to the shallow trap, $E_b$.  The heavily compensated nature of
the system gives rise to random potentials coming from the charged
centers responsible for compensation. We model this through an on-site
random potential $u(i)$ which leads to the second term in
Eq. \ref{1}. A full model would require a self-consistent
determination of this potential taking in consideration screening
processes.  The third term is the AFM interaction between Mn spins
$\vec{S}(i)$ and charge carrier spins. Since the Mn spins are very
localized (the $3d$ shell has a radius of the order 1-2\AA) the
antiferromagnetic exchange integral is taken to be of the form $J_{ij}
=J \vert \phi(i,j) \vert^2= J \exp{\left(-2 { \vert \vec{R}_i -
\vec{R}_j \vert / a_B} \right) }$, reflecting the probability of
finding the charge carrier in the $s$-type shallow level about
impurity $j$ near the $i$th Mn spin.  In this notation, $J$ is simply
the AFM exchange integral of an isolated localized hole centered at a
Mn impurity with the spin of the Mn. The last line in Eq. (\ref{1})
describes the interaction of the carrier and Mn spins with an external
magnetic field H.

\subsection{ Mean-Field Approximation  }

We treat the antiferromagnetic interaction at the mean-field level,
given by the factorization
\begin{equation}
\label{3}
\vec{S}(i)\cdot \hat{\vec{\sigma}}_j \rightarrow S_{Mn}(i)
\hat{\sigma^z_j} + S^z(i) s_h(j) - S_{Mn}(i)s_h(j),
\end{equation}
where $S_{Mn}(i) = \langle S^z(i) \rangle$ and $s_h(j) = \langle {\hat
\sigma}^z_j \rangle $ are the expectation values of the $i$th Mn spin
and of the total spin created by charge carriers at the $j$th Mn site,
respectively, and must be computed self-consistently at each site.
For simplicity of notation, we used $\hat{\vec{\sigma}}_j =
c^{\dagger}_{j\alpha} {1 \over 2} \vec{\sigma}_{\alpha\beta}
c_{j\beta}$ in Eq. (\ref{3}). In writing Eq. (\ref{3}) we made the
implicit assumption that the rotational symmetry is broken in the
direction of the external magnetic field H (defining the z-axis) for
all spins. In fact, one may start with a Heisenberg-like factorization
in the absence of external magnetic fields, but we find that in the
self-consistent configurations the spins are always collinear, and as
a result we regain the Ising-like factorization of Eq. (\ref{3}).
Disorder induced non-collinear ground-states have been recently
suggested in a model which assumes that holes occupy a Fermi sea in
the valence band. \cite{MacDSch}

The mean-field Hamiltonian can be separated into three parts
\begin{equation}
\label{4}
{\cal H}_{MF}= {\cal H}_{spin} + {\cal H}_{cc}+ {\cal H}_{const}.
\end{equation}

The spin Hamiltonian may be rewritten as
\begin{equation}
\label{5}
{\cal H}_{spin} = - \sum_{i}^{} H_i S^z(i),
\end{equation}
where
\begin{equation}
\label{6}
H_i = \tilde{g} \mu_B H - J \sum_{j}^{}\vert \phi(i,j)\vert^2 s_h(j)
\end{equation}
is the effective magnetic field for the Mn spin at site $i$, including
the charge carrier contributions. Then, the average Mn spin at site
$i$ is
\begin{equation}
\label{7}
S_{Mn} (i) = {\cal B}_S(\beta H_i),
\end{equation}
where ${\cal B}_S(x)= (S + {1 \over 2}) \coth{[(S+{1\over 2})x]}-{ 1
 \over 2} \coth{{x\over 2}}$ is the Brillouin function corresponding
 to the spin $S=5/2$, and $\beta=1/k_BT$.  The contribution to the
 internal energy from the spin Hamiltonian equals
\begin{equation}
\label{6.1}
U_{spin}(T)=-\sum_{i}^{}H_iS_{Mn}(i).
\end{equation}

The charge carrier Hamiltonian can be rewritten as
\begin{equation}
\label{8}
{\cal H}_{cc} = \sum_{i,j}^{}t_{ij} c^{\dagger}_{i\sigma} c_{j\sigma}
+\sum_{i\sigma}^{}\left(\epsilon_{i\sigma} - \mu\right)
c^{\dagger}_{i\sigma} c_{i\sigma},
\end{equation}
where
\begin{equation}
\label{9}
\epsilon_{i\sigma} = { \sigma \over 2} \left( J \sum_{j}^{}\vert
\phi(i,j) \vert^2 S_{Mn}(j) - g \mu_B H\right) + u(i)
\end{equation}
is the effective on-site energy created by the Mn spins, the on-site
disorder and the external magnetic field. A chemical potential $\mu$
has been added since we treat the charge carriers in the
grand-canonical ensemble.

The charge carrier Hamiltonian can be diagonalized to obtain:
\begin{equation}
\label{10}
{\cal H}_{cc} = \sum_{n\sigma} (E_{n\sigma} - \mu)
a^{\dagger}_{n\sigma} a_{n\sigma},
\end{equation}
using the linear combinations
\begin{equation}
\label{11}
a^{\dagger}_{n\sigma} = \sum_{i}^{} \psi_{n\sigma}(i)
c^{\dagger}_{i\sigma}, \hspace{10mm}
\end{equation}
where $\psi_{n\sigma}(i)$ is the probability amplitude to find a
charge carrier occupying level $n$ with spin $\sigma$ in the shallow
state centered at site $i$. The chemical potential is given by the
condition
\begin{equation}
\label{11b}
N_h = \sum_{n\sigma}^{} f(E_{n\sigma}),
\end{equation}
where $f(E_{n\sigma}) = \left[\exp{\left(\beta(E_{n\sigma}-
\mu)\right)} + 1\right]^{-1}$ is the Fermi distribution describing the
occupation probability of the level $(n\sigma)$.

The expectation value of various charge carrier operators can be
computed in a straightforward way.  In particular, the average spin
created by charge carriers at site $i$ is
\begin{equation}
\label{12}
s_h(i) = { 1 \over 2} \sum_{n}^{} \left[
\vert\psi_{n\uparrow}(i)\vert^2 f(E_{n\uparrow}) -
\vert\psi_{n\downarrow}(i)\vert^2 f(E_{n\downarrow}) \right],
\end{equation}
while the contribution of the charge carriers to the total internal
energy is
\begin{equation}
\label{12.1}
U_{cc}(T) = \sum_{n\sigma}^{} E_{n\sigma} f(E_{n\sigma}).
\end{equation}

Finally, the third term in Eq. (\ref{4}) contains the constant terms
from the mean-field factorization
\begin{equation}
\label{12.3}
{\cal H}_{const}=-J\sum_{i,j} \vert \phi(i,j)\vert^2 S_{Mn}(i)s_h(j)
  =U_{const}(T).
\end{equation}
As a result, the total internal energy of the system is given by
\begin{equation}
\label{12.2}
U(T) = - \tilde{g} \mu_B H \sum_{i}^{} S_{Mn}(i) + \sum_{n\sigma}^{}
 E_{n\sigma} f(E_{n\sigma}),
\end{equation}
where, in the absence of external magnetic fields, all the
contribution comes from $U_{cc}(T)$. However, the charge carrier
Hamiltonian ${\cal H}_{cc}$ contains the interaction with the average
Mn spins incorporated in the eigenenergies $E_{n\sigma}$, so in fact
their contribution is also included in this term.

The specific heat of the entire system is given by
$$
C_V(T)={\partial U(T) \over \partial T}.
$$
Various other quantities of interest can be computed in a similar
fashion.

We solve the mean-field equations (\ref{6})-(\ref{12}) using an
iterative algorithm. We start with a guess for the initial $S_{Mn}(i)$
configuration for each temperature $T$ of interest (details about this
are provided in section III.B). We numerically diagonalize the charge
carrier Hamiltonian Eq. (\ref{8}) and find the charge carrier
eigenvalues, eigenfunctions as well as the chemical potential from
Eq. (\ref{11b}). This allows us to compute the expectation values for
the charge carrier spins at each site $s_h(j)$ [from Eq. (\ref{12})]
and therefore obtain the effective magnetic field at each Mn site
$H_i$ [Eq. (\ref{6})]. The new expectation values for the Mn spins at
each site $S_{Mn}(i)$ are then obtained from Eq. (\ref{7}) and the
iterations are repeated until self-consistency is achieved. We define
self-consistency as corresponding to the situation where the largest
absolute variation of the on-site charge carrier energies
$\epsilon_{i\sigma}$ [Eq. (\ref{9})] between two successive iterations
is less than $10^{-3}$. This corresponds to relative errors of
$10^{-5}$ or less for all computed quantities.

\subsection{Parameters}

We consider a $N\times N\times N$ FCC sublattice of the valence-III
element, of lattice constant $a$ ($a=5.65\AA$ for GaAs). We assume
throughout this paper that the system has periodic boundary
conditions. The Mn doping is characterized by $x= N_d/4N^3$, leading
to a Mn concentration $c_{Mn} = 4x/a^3$. The hole concentration is
$c_h=p c_{Mn}$, where $p = N_h/N_d$. Values of interest are
$x=0.01-0.05$ and $p=5-10\%$.\cite{Besch} The choice of the system
size $N$ is described in the following subsection.

Throughout this paper we use parameters specific to the
Ga$_{1-x}$Mn$_x$As system. The binding energy of the hole (defining
the Ry unit of this problem) is $E_b=112.4$ meV=1Ry. \cite{BG} Using
the Luttinger Hamiltonian in the spherical approximation\cite{BG} we
find the effective mass for the heavy hole to be
\begin{equation}
\label{3.1}
m_h = { m_e \over \gamma_1 - (6\gamma_3+4\gamma_2)/5} = 0.56~m_e
\end{equation}
where the $\gamma$-coefficients for GaAs are $\gamma_1=7.65$,
$\gamma_2=2.41$ and $\gamma_3=3.28$.\cite{BG} This allows us to
estimate the Bohr radius of the isolated impurity state as\cite{SE}
\begin{equation}
\label{3.2}
a_B= { \hbar\over\sqrt{(2m_h E_b)} } = 7.8~\AA.
\end{equation}
This is in excellent agreement with another possible estimate,
obtained by assuming that the hole is bound to its impurity by a pure
Coulomb attraction, in which case $E_b = e^2/(2\epsilon a_B)$. Using
the value $\epsilon=10.66$ for GaAs leads to $a_B=7.82\AA$. While this
agreement is probably fortuitous, similar values have been used in
literature.\cite{Esch} The characteristic value of the hopping
integral is $t(4a_B)=20$meV.  Besides the binding energy $E_b$, a
second energy scale is provided by the exchange integral. We use the
value $J= 3 \epsilon= 15$ meV, where $\epsilon= 5$ meV is the value
obtained in Ref. 7  for the antiferromagnetic interaction of
an isolated hole with the spin of its own trapping Mn impurity. We
include the factor of 3 as the simplest way to account for the fact
that the heavy holes have spin projections $j_z = \pm {3 \over 2}$,
while in our model they are modeled as $s_z= \pm {1 \over 2}$
objects. The final parameter, $W$, (or its dimensionless counterpart
$W/E_b$) characterizes the on-site disorder due to uncompensated
impurities. We assume that $u(i)$ has a uniform distribution in a
range $[-W, W]$. Following Ref. 8  we find an estimate of
$W$ given by $e^2/\epsilon \tilde{r}$, where $\tilde{r} \sim
1/n_{Mn}^3$ is roughly the average distance between Mn impurities. For
the typical charge carrier concentration considered $c_h= 1.5\times
10^{20}$ cm$^{-3}$ and $p=10\%$, we find $\tilde{r} \sim 9\AA \sim
a_B$, which suggests $W \sim$ 1 Ry. As we show later, while the
magnitude of on-site disorder $W$ influences the shape of the
magnetization curve, the critical temperature $T_c$ has only a very
weak dependence on it. As a result, and given the fact that the
compensation mechanism in these systems has not yet been fully
clarified, we do not attempt a more detailed modeling of the on-site
disorder at the present stage.
 
Thus, in the absence of external magnetic fields, the problem depends
on five dimensionless parameters, $J/E_b$, $a_B/a$, $n_ha_B^3$, $x$
and $W/E_b$.

\section{Results}

\subsection{Simple cubic superlattice of Mn}

In order to gain some insight in the behavior of the system, we first
consider ths simplified case of Mn impurities placed in a simple cubic
structure, with a superlattice constant $a_L = a / (4x)^{1/3}$.
Strictly speaking, only concentrations $x$ for which $a_L$ is
commensurate with $a$ would be physically acceptable.  However, since
we perform this calculation only to get a feeling for the homogeneous
solution, we disregard the underlying GaAs lattice in this particular
case and assume that Mn spins could be located anywhere in space. We
also set the on-site disorder $u(i)=0$, and turn off the external
magnetic field ($H=0$).

For the ordered case, translational symmetry implies $s_h(i)=s_{hole},
S_{Mn}(i) = S_{Mn}$ for all sites $i$ of the cubic Mn superlattice. As
a result, the charge carrier Hamiltonian ${\cal H}_{cc}$ is
diagonalized with plane waves, and the self-consistent equations
(\ref{6})-(\ref{12}) reduce to:
\begin{equation}
\label{20}
S_{Mn} = - {\cal B}(\beta J_{eff} s_{hole}),
\end{equation}
\begin{equation}
\label{21}
s_{hole} = { 1 \over 16 \pi^3} \int d\vec{k} \sum_{\sigma}^{} \sigma
f(E_{\vec{k}\sigma}),
\end{equation}
where the chemical potential is determined from
\begin{equation}
\label{22}
p = {N_h \over N_d }= { 1 \over 8 \pi^3} \int d\vec{k}
\sum_{\sigma}^{} f(E_{\vec{k}\sigma})
\end{equation}
and the charge carrier eigenenergies are given by
\begin{equation}
\label{23}
E_{\vec{k}\sigma} = \epsilon(\vec{k}) + {\sigma \over 2} J_{eff}
S_{Mn}.
\end{equation}
Here, $\vec{k}$ is measured in units of $1/a_L$, and the integrals in
Eqs. (\ref{21}) and (\ref{22}) are performed over the first Brillouin
zone $-\pi < k_{\alpha} \le \pi$, $\alpha=x,y,z$. The non-interacting
charge carrier dispersion relation is given by
\begin{equation}
\label{25}
\epsilon(\vec{k}) = \sum_{i \ne 0 }^{} t(r_i) e^{i \vec{r}_i \cdot
\vec{k}}
\end{equation}
and
\begin{equation}
\label{24}
J_{eff} = J \sum_{i}^{} |\phi(i)|^2 = J \sum_{i}^{} e^{- {2 r_i \over
a_B}}.
\end{equation}
In Eq. (\ref{25}) and (\ref{24}) the sums are performed over the whole
crystal, but the exponential decay of $|\phi(i)|^2 $ and $t(r_i)$ lead
to finite results.

The self-consistent values of $0< S_{Mn}< 2.5 $ and the average charge
carrier spin $-0.5< s_h = \sum_{i}^{} s_h(i) / N_h = s_{hole} N_d/N_h
<0 $ obtained for $x=0.01-0.05$ and $p=10\%$ are shown in
Fig.\ref{fig1}. The overall signs indicate the AFM alignment of the
charge carrier and Mn spins.

The total magnetization of the sample, obtained by adding the Mn and
charge carrier contributions, looks similar to the $S_{Mn}$ curve,
since the number of Mn spins is $1/p\sim 10$ times larger than the
number of charge carrier spins, and they also have larger
g-factors. Thus, we see that the magnetization curve does not have the
typical form of ferromagnetic systems.  In fact, each curve shows
three different regimes. Below $T_c$ there is a region where neither
the charge carrier nor the Mn spins are yet fully polarized. The gap
$J_{eff} S_{Mn}$ between the $\sigma=\downarrow$ and $\sigma=\uparrow$
charge carrier bands (see Eq. (\ref{23})) increases quickly as the
temperature decreases. This leads to a polarization of the charge
carriers, which in turn polarizes the Mn spins even more. Since the
number of charge carriers is relatively small, they are the first to
fully polarize, at a characteristic temperature defined by
$J_{eff}S_{Mn} - E_F \sim k_B T$. Here, $E_F$ is the Fermi energy of
charge carriers measured from the bottom of the $\sigma=\downarrow$
band, and the condition simply means that the gap from the highest
occupied $\sigma=\downarrow$ level to the first available
$\sigma=\uparrow$ level is larger than the thermal energy. As a
result, below this temperature charge carriers are fully
spin-polarized, $s_{hole}=ps_h=-0.5p$. From Eq. (\ref{21}) we see that
below this temperature, the Mn spins behave as if they are in a
constant external magnetic field of magnitude $H=J_{eff}s_{hole}$. For
temperatures such that $\beta H \ll 1$, the Brillouin function may be
linearized and we find that $S_{Mn} \sim J_{eff}s/k_BT$ (the Curie
law), and $S_{Mn}$ increases roughly like $1/T$ as the temperature
decreases. This explains the uncharacteristic concave upward shape of
the Mn spin magnetization. Finally, below temperatures for which
${\cal B}_{S}(\beta J_{eff}s) \approx S $ (i.e. $k_BT < 3p
\sum_{i}^{}e^{-{2 r_i\over a_B}}$ meV), the Mn spins are also fully
polarized.

In the inset of Fig.\ref{fig1} we plot the specific heat per Mn
impurity for the same parameters. In all cases we see two distinct
contributions. The lower peak is entirely due to the Mn spins, while
the upper one is the charge carrier contribution. At low temperatures
the charge carriers are all ``frozen'' at the bottom of the
$\sigma=\downarrow$ band, and all the entropy is due to fluctuations
in the Mn spins. This can be easily checked by computing $C_V^{spins}=
dU_{spin}/dT$, with $s_{hole}=-0.5p$ substituted in
Eq. (\ref{6.1}). This accounts for the entire lower peak. At the
higher temperatures the Mn spins are almost free (the effective
magnetic field orienting them is very small), and therefore right
below $T_c$ the entropy is dominated by spin fluctuations of the
charge carriers.

For increasing charge carrier concentrations $p$, one expects that the
temperature where charge carriers become fully polarized decreases
(since $E_F$ increases) and thus the unusual regime with fully
polarized charge carriers and $S_{Mn} \sim 1/k_B T$ is restricted to
smaller intervals. In other words, we expect that for larger $p$
values the Mn spin magnetization should begin to look more like the
characteristic convex upward (concave downward) form seen in usual
ferromagnetic systems.  This is confirmed in Fig. \ref{fig2}, where
the average charge carrier and Mn spins are plotted for $x=0.02$ and
$p=5,10,25$ and $40\%$.

It is interesting to note that the critical temperatures obtained for
this homogeneous case using the nominal parameters and Bohr radii from
the literature are actually in good agreement with experimentally
measured values. In Fig. \ref{fig3} we show the critical temperatures
for three GaMnAs samples,\cite{Besch} which are estimated to have
$p=5-10\%$. The experimental points fall right in between the
theoretical curves corresponding to the two charge carrier
concentrations $p$. However, it is important to emphasize the fact
that fairly small variations in any of the parameters can lead to
rather large variations in $T_c$. Indeed, from Fig. \ref{fig2} we see
how sensitive $T_c$ and the shape of the magnetization are to $p$.  If
we increase the Bohr radius by just 1$\AA$, while keeping all the
other parameters fixed, the critical temperatures increase by roughly
$50\%$, and the experimental points are well below the new $p=5\%$
theoretical estimations.  In the following section we show that
disorder in Mn positions has, at least at the mean-field level, a
large effect on $T_c$. At the same time, the mean-field approximation
itself underestimates thermal fluctuations, and therefore may
substantially overestimate critical temperatures. Finally, the actual
transition temperature varies substantially with the precise form for
the hopping parameter (see section F). Therefore, we conclude that the
good agreement shown in Fig. \ref{fig3} is likely fortuitous.

\subsection{Effects of disorder in Mn positions}

We now analyze the effects of randomness in the Mn impurity positions
on the shape of the magnetization curve and the value of the critical
temperature. We restrict ourselves in this section to the case of no
on-site disorder, i.e. $u(i)=0$, and vanishing external magnetic
field. In this case, the system is no longer homogeneous. As a result,
we have to limit ourselves to a finite $N \times N\times N$ FCC
lattice with periodic boundary conditions, choose randomly the
positions of the $N_d$ Mn spins and solve Eqs.(\ref{6})-(\ref{12})
self-consistently for each site.

A technical question is how large a system we should consider in order
to assume that the ``bulk limit'' has been reached. The size $N$ of
the system is chosen so as to minimize finite size effects. These are
monitored through both their effect on the magnetization curves
(especially on $T_c$), as well as on the total density of states
(DOS). We find that for $x=0.05$, finite-size effects become
negligible for systems which have more than $N_h\sim 50$ holes. As a
result, the minimum size we use for this concentration is $N=14$,
corresponding to $N_d=548, N_h=55$. For the smallest Mn concentration
we investigated, $x=0.00926$, we find that finite size effects are
small for systems having as few as 12 holes (corresponding to $N=15$,
close to the previous value). However, even for this lower
concentration, most of the results shown are obtained for lattices of
size $N=24$, corresponding to $ N_d=512, N_h=51$. This ensures that
grand-canonical fluctuations in the total number of charge carriers
are minimized as well.

To solve the equations (\ref{6})-(\ref{12}) self-consistently, we
start with different initial Mn spin values $S_{Mn}(i)$ to use in the
charge carrier Hamiltonian ${\cal H}_{cc}$ for the first iteration of
the process to self-consistency. We first consider biased initial
conditions. In this case, we start the first iteration for the lowest
temperature considered by assuming that all Mn spins are fully
polarized, $S_{Mn}(i)=2.5$. After several iterations, the
self-consistent values $S_{Mn}(i)$ corresponding to this temperature
are found. We then use these values as the initial guess for the next
higher temperature considered, etc. This allows us to find, at each
temperature, the self-consistent solution with the highest possible
total magnetization.  We next start each search with {\em random
values} for both the magnitude and the sign of $S_{Mn}(i)$ for each
temperature considered.  Since in principle there could be metastable
solutions, the ``true'' mean-field solution is the one corresponding
to the lowest total free energy.

In Fig.\ref{fig5}, left panel, we show the expectation values for the
 average Mn spin $S_{Mn} = 1/N_d \sum_{i} S_{Mn}(i)$ and the average
 charge carrier spin $s_h=1/N_h \sum_{i} s_{h}(i)$ obtained for one
 disorder realization using biased (full lines) and random initial
 configurations (circles) for $p=10\%$ and $x=0.00926$. For random
 initial conditions we actually plot $|S_{Mn}| > 0$ (full circles) and
 $-|s_h|<0$ (empty circles), since the two expectation values always
 have opposite sign, but the orientation is arbitrary for random
 initial conditions. Random initial conditions lead to various
 self-consistent configurations, with magnetizations smaller or equal
 to the maximum possible value given by the biased configuration. The
 smaller value of average magnetizations for the random initial
 conditions configurations is not a consequence of smaller
 polarizations of individual spins, but of the appearance of regions
 with local magnetizations pointing in different directions.  The
 right panel of Fig.\ref{fig5} shows the difference between the total
 energy per Mn spin obtained with random initial conditions, and that
 of the biased configurations.  At low temperatures, all the random
 configurations are much higher in energy than the biased
 configuration, with the difference increasing for configurations with
 lower total magnetization. However, by $k_BT/J \sim 0.5$ the energy
 difference between configurations becomes comparable to $k_BT$, as
 evident in the intersection with the solid line which is a plot of
 $k_BT/N_d$. At these temperatures thermal fluctuations will enable
 the system to vary continuously among these various states,
 effectively suppressing the magnetization and therefore lowering
 $T_c$.  A proper treatment of the effect of thermal fluctuations
 requires going beyond the mean-field approximation, e.g. by a Monte
 Carlo simulation.

We have found similar results for various other Mn and hole
densities. We have looked in all at over 150 samples of varying sizes
$N=14-24$ corresponding to Mn concentrations $x=0.00926-0.05$ and
relative hole concentrations $p=10-30\%$. In all cases, the most
polarized (biased) state has been found to have the lowest total free
energy in our model. {\em For larger $x$, we in fact find that most
random initial samples converge to the biased limit, signifying a more
robust magnetization than at lower $x$}.  We conclude that for this
range of concentrations the system is indeed ferromagnetic at low
temperatures, and that the biased configuration curves may be used to
obtain the lowest energy mean-field configuration possible. However,
the $T_c$ given by this mean-field biased curve is likely to be
significantly larger than the one provided by a Monte-Carlo
simulation, with the difference likely to grow as the concentration
decreases.

Comparing typical $T_c$ values obtained for random Mn configurations
(Fig. \ref{fig5}) with those obtained for the ordered Mn lattices with
similar $x$ and $p$ values (see Fig. \ref{fig1}) we see that the shape
of the magnetization curves is significantly changed by
randomness. $T_c$ is increased, while the curves become even more
concave. In fact, the Mn spins do not reach the saturation limit
$S_{Mn}=2.5$ until very low temperatures. While at first sight this
significant increase of $T_c$ for the disordered case may seem
puzzling, in fact it has a very physical explanation. In the
disordered sample there are regions of higher local concentration of
Mn. The charge carriers prefer these regions, since they can lower
their kinetic energy by moving among several nearby Mn sites. They can
also lower their magnetic energy by polarizing the spins of these Mn
impurities. As a result, these regions of higher Mn concentration will
become spin-polarized at higher temperatures than the average sample,
pushing $T_c$ up.

This point can be illustrated by looking at a histogram of the total
density of charge carriers at site $i$, defined as
\begin{equation}
\label{4.1}
\rho(i) = \sum_{j}^{}\vert \phi(i,j)\vert^2 \langle
c^{\dagger}_{j\uparrow} c_{j\uparrow} + c^{\dagger}_{j\downarrow}
c_{j\downarrow}\rangle.
\end{equation}
[In other words, $\rho(i)$ is given by the probability $ p(i)=\langle
c^{\dagger}_{i \uparrow} c_{i\uparrow} + c^{\dagger}_{i\downarrow}
c_{i\downarrow}\rangle $ of finding a charge carrier at site $i$, plus
exponentially small contributions due to tailing from charge carriers
found on nearby Mn sites].  Since at low temperatures only spin down
charge carrier states are significantly occupied, we can also
interpret $\rho(i)$ as being proportional to the effective magnetic
field acting on the $i$th Mn spins [see Eq. (\ref{6})], with external
magnetic field $H=0$). In Fig. \ref{fig6}a we show, on a logarithmic
plot, such a histogram obtained for 25 random Mn configurations with
$x=0.00926$ and $p=10\%$ at $k_BT/J=0.0006$ (dotted line) and
$k_BT/J=0.6$ (full line).  The vertical line indicates the position of
the $\delta$-function for the histogram corresponding to the
homogeneous system (ordered Mn superlattice).  In that case the
density $\rho(i)$ at every site is the same, and is given by $\rho = p
\sum_{i}^{}\exp{(-2r_i/a_B)}$. For small concentrations $x$ the sum is
roughly equal to unity, and the charge carrier density and effective
magnetic field at each Mn site is $p$, respectively $Jp$.  On the
other hand, for the random (disordered) configurations, a
double-peaked structure is clearly visible. There is one sharp peak
centered about $\rho \approx 0.6 > p$ corresponding to densities much
higher than the average, and a second much broader peak centered at
exponentially small values $\rho \approx 10^{-2} \ll p $.  In other
words, there are some Mn sites with a large charge carrier density,
which strongly polarizes the respective Mn spins up to high
temperatures. These correspond to the high density regions of Mn, and
define a $T_c$ much higher than the one of the homogeneous system.
The rest of the Mn sites have a small charge carrier density (or
effective magnetic field) and therefore they very quickly become
depolarized as the temperature increases, leading to the fast decrease
in the average Mn spin value $S_{Mn}$. This phenomenology can be
captured quite accurately by dividing the spins into strongly and
weakly interacting ones, depending on whether their effective magnetic
field is larger or smaller than the corresponding thermal
energy.\cite{Malcolm} A mix of ferromagnetic and paramagnetic
contributions to the $M(H)$ curves, which can be attributed to
strongly, respectively weakly interacting spins, has also been
observed experimentally. \cite{oiwa} As the temperature increases to
$k_BT/J=0.6$ (just below $T_c$ for this density), the double-peaked
structure is still apparent, although it becomes more centered around
the average value and the peak corresponding to strongly interacting
spins decreases. For temperatures well above $T_c$ (see inset of
Fig. \ref{fig6}a) the distribution width decreases even more, although
it still extends over more than two orders of magnitude. This behavior
suggests that as the temperature is lowered through $T_c$, the charge
carriers start to polarize the most dense clusters of Mn. As a result,
their wave-functions become more concentrated in these high-density Mn
areas, where the carriers can further lower their magnetic exchange
energy. This leads to the increased weight of the higher $\rho(i)$
peak. The concentration of charge carriers in the high-density areas
implies a further depopulation of the low Mn density areas, pushing
the low edge of the $\rho(i)$ distribution towards lower values. With
decreasing temperature the histogram quickly changes and for $k_BT/J <
0.3$ it already has a shape identical to the that of the low
temperature $k_BT/J=0.0006$ histogram.

In Fig. \ref{fig6}b we compare the low temperature histograms for
different concentrations $x=0.00926$ (dotted lines) and $x=0.05$ (full
lines). The vertical lines again show the corresponding values for the
homogeneous (ordered) systems. For $x=0.05$ the double-peak structure
is also clearly visible. However, the whole histogram shifts to higher
$\rho(i)$ values.  This is consistent with the fact that for larger
charge carrier concentrations the overall interactions are increased.
[It is interesting that for $x=0.05$ there is a finite concentration
of sites for which $\rho(i) > 1$ ($\log_{10}[\rho(i)]>0$). Since the
probability of finding a hole at any site $ p(i)=\langle
c^{\dagger}_{i \uparrow} c_{i\uparrow} + c^{\dagger}_{i\downarrow}
c_{i\downarrow}\rangle <1 $, this suggests that in this case some Mn
spins strongly interact with several charge carriers that are nearby
[see Eq. (\ref{4.1})]. For $x=0.00926$, however, $\rho(i) < 1$ for all
sites, suggesting that Mn spins interact at most with one charge
carrier].

To further check this picture of enhancement of $T_C$ within our
model, we have ``tuned'' the amount of disorder (randomness) in the Mn
positions.  Fig. \ref{fig7} shows curves of average charge carrier and
Mn spins as a function of temperature for four different distributions
of Mn spins. They all correspond to the same values $x=0.00926$ and
$p=10\%$. The curve with the lowest $T_c$ is the curve for the ordered
Mn superlattice, with $a_L=3a$. The next curve corresponds to a weak
disorder configuration in which each Mn atom is allowed to randomly
choose one of the 12 nearest neighbors of the underlying FCC
sublattice. In other words some randomness has been allowed for,
although the configuration is still quite homogeneous, with one Mn
site inside each cubic supercell. Even this small amount of disorder
is seen to have a significant effect on the shape of the magnetization
and the mean-field $T_c$ value. Some Mn spins now have a higher
effective charge carrier concentration, pushing $T_c$ to higher
values. However, at low temperatures we see that the average Mn spin
is smaller than that of the ordered lattice, meaning that some Mn
spins are in much lower effective magnetic fields (have lower local
charge carrier density) and only get saturated at much lower
temperatures.  The third curve corresponds to a medium disorder
configuration in which Mn impurities are allowed to pick any sites on
the FCC sublattice, as long as the distance between any two of them is
at least $2a$. This allows for even more randomness, leading to even
higher $T_c$, while at low temperatures the average Mn spin is even
more suppressed. Finally, the curve with the highest $T_c$ corresponds
to a completely random (strong disorder) Mn configuration. Monitoring
the histogram of on-site densities for these cases, we find the
expected behavior: increasing disorder leads to a larger spread of the
densities about the average value of the ordered superlattice, leading
to both the increase of mean-field $T_c$ as well as the decrease of
the saturation temperature.

A qualitatively similar picture holds for higher Mn concentrations as
well as higher hole densities, although the effects are quantitatively
substantially less. In Fig. \ref{fig8} we show the Mn and hole spins
for both the simple cubic superlattice and the random Mn distribution
on the FCC sublattice for $x=0.05$ for two different $p$.  While $T_C$
is again substantially larger in the random system, the percentage
increase is smaller than in the $x=0.00926$ case. Increasing the hole
concentration from $p=10\%$ to $p=30\%$ makes the curve shape more
conventional (convex upward).  The reason is simply that the
fluctuations in the local doping are smaller at higher Mn
concentrations, and increased hole doping further reduces the width of
the density distribution.

All the curves shown so far correspond to one particular random
distribution of the Mn impurities on the FCC sublattice. As the amount
of disorder in the positions of the Mn sites increases, so do the
variations between curves corresponding to various realizations of
disorder in this mean-field approach. In Fig. \ref{fig9} we show a
typical spread for 25 different disorder realizations for both
$x=0.00926$ (upper panel) and $x=0.05$ (lower panel). In both cases we
see that while the curves have similar shapes, there is a significant
variation present. In particular, for the lower concentration, we see
that most magnetization curves have a very elongated tail near $T_c$,
arising from rare dense clusters of (usually nearest neighbor) Mn
spins that are polarized by the holes. This tail will likely be
destroyed by thermal fluctuations if one goes beyond the mean-field
treatment. For the higher concentration most curves have $k_B T_c \sim
0.9 J$, although again there are a few which have longer tails arising
from magnetization of dense clusters of Mn impurities.

\subsection{Metal-Insulator Transition}

According to Mott's criterion, a doped semiconductor goes through a
metal-insulator transition (MIT) for a charge carrier concentration
given by $n_h^{1/3}a_B \sim 0.25$. (For compensated systems, the
critical density is experimentally found to be somewhat larger).
Neglecting the effects of compensation and assuming $p=10\%$, for
$a_B=7.8\AA$ this corresponds to a Mn concentration $x\sim
0.015$. Given the variations in the compensation parameter $p$ for
different samples, this is in good agreement with experimental
measurements which indicate a MIT at $x\sim 0.03$.\cite{Matsu}

One way of monitoring the MIT is by determining whether the charge
carrier states in the vicinity of the Fermi level are localized or
extended. We characterize the charge carrier states using the Inverse
Participation Ratio (IPR), defined for each state $(n\sigma)$ by
$$
IPR(n\sigma) = { \sum_{i}^{} |\psi_{n\sigma}(i)|^4 \over
\left(\sum_{i}^{} |\psi_{n\sigma}(i)|^2\right)^2}.
$$
For a state extended over the $N_d$ sites of the system, one expects
$|\psi_{n\sigma}(i) | \sim 1/\sqrt{N_d}$, and therefore $IPR(n\sigma)
\sim 1/N_d$. In other words, for extended states $IPR(n\sigma)$ is
inversely proportional to the size of the system. For localized
states, $IPR(n\sigma)$ is inversely proportional to the number of
sites over which the wavefunction is localized, and therefore
independent of the size of the system.

In Fig. \ref{fig10} we plot $IPR(E,\sigma)$ for a completely
disordered system with $x=0.00926$ and $p=10\%$, at a low temperature
($k_BT/J=0.0006$).  For each system size we consider 100 different
disorder realizations, and we average the $IPR$ of all the states with
eigenvalues within 2 meV of each other. We show both the
$\sigma=\downarrow$ and $\sigma=\uparrow$ sub-bands for three system
sizes, $N_d=125, 512 $ and 1000, corresponding to $N_h=12, 51$ and
100. A large gap of size $4.2J = 63$ meV opens between the two
subbands, and only states in the $\sigma=\downarrow$ are occupied at
this low temperature. The position of the Fermi energy is shown by the
vertical line, at about 45 meV above the bottom of the band. We
clearly see that states at the bottom of either band are localized,
with the $IPR$ independent of the system size. At higher energy,
however, the states become extended, with the $IPR \sim 1/N_d$. The
mobility edge for the $\sigma=\downarrow$ band is at an energy of
about -100 meV. The Fermi level is below the mobility edge, signifying
that this system is still insulating, in agreement with experimental
measurements. However, the IPR of the states at the Fermi energy is
less than a factor of 2 larger than the IPR at the mobility edge for
our sizes. This suggests that significant tunneling may occur in
between various regions occupied by the holes, leading to alignment of
polarization of all the high-density regions.  The inset shows the
density of states for the two spin-polarized subbands. The curves
corresponding to the three system sizes fall on top of each other,
proving that the ``bulk limit'' is already reached. In all cases we
investigated, the DOS has an extremely long upper tail, only part of
which is shown. Due to the small relative concentration of holes
$p=10\%$, only states very close to the bottom of the impurity band
are occupied. As already mentioned, this supports our assumption that
neglecting band states (which are roughly 110 meV above the impurity
band) is a good starting approximation.

In Fig. \ref{fig11} we show the IPR for the $\sigma=\downarrow$
subbands of configurations with medium disorder (satisfying the
restriction that the distance between any two impurity sites is larger
than $2a$). Again, 100 configurations each for system of size
$N_d=125,512$ and 1000 are averaged. The Fermi energy of the
medium-disorder system is shown by the dashed line. The
$\sigma=\uparrow$ sub-band (which is completely empty) is not shown.
For the medium-disorder system the IPR values are lower than those
corresponding to the strongly disordered system. This is the expected
behavior, since in the limit of no disorder all wave functions must
become fully extended and the system is metallic. In fact, we observe
that even for the medium disorder case the Fermi energy is just above
the mobility edge.

The histograms presented in Fig. {\ref{fig6}a show that as the
temperature increases the distribution of charge carriers in the
system becomes somewhat more homogeneous (the width of the
distributions decreases). This suggests that the charge-carrier
wave-functions become more extended at higher temperatures, and
therefore the system is more ``metallic''. This is in qualitative
agreement with resistivity measurements\cite{Matsu} which show a
larger resistivity at low temperatures than above $T_c$ for all
low-density samples.  We conclude that while the exact value of $p$
and nature of disorder in Mn positions play a crucial role, the
$x=0.00926$ sample is most likely insulating.

IPR curves for $x=0.05$ and completely random samples (strong
disorder) are shown in Fig. \ref{fig12}. In this case, we have used
larger size systems with $N_d=548, 981$ and 1600 in order to avoid
finite size effects.  Again, we see that states at the bottom of
either subband are localized, while states at higher energies are
extended (their IPR scales with $1/N_d$). In this case, the system is
clearly above the metal-insulator transition, in qualitative agreement
with experiment.

Experimentally it has also been observed \cite{Esch,Katsu,Matsu} that
samples with even higher concentrations ($x> 0.07$) become insulating
again. However, these samples also seem to have a much lower relative
charge carrier density $p$ (see Ref. 18).   Since the
$x=0.05$ system is just above the MIT, it is reasonable to assume that
a significant decrease in the density of charge carriers (leading to a
significant decrease in the Fermi energy) could move the Fermi level
below the mobility edge and therefore be responsible for the
re-entrant insulating state.

The fact that these systems are either metallic, or not too far from
the MIT, is important for obtaining a large critical temperature.
Delocalization of the electronic wavefunction over several sites,
which allows the Mn spins to effectively communicate with each other,
is the essential ingredient which leads to alignment of the
polarization of all the high density regions. The charge carriers
hopping between various high-density areas will force the alignment of
Mn spins in each region to be the same, in order to lower their
kinetic energy. However, maximizing the critical temperature seems to
require a fine balance: increasing the disorder leads to increased
$T_c$, but also to increased localization. If the charge carrier
states become so localized that there is no tunneling in between high
density occupied regions, than the direction of polarization of each
such region is uncorrelated with the direction of polarization of the
other regions, and the average magnetization of the sample will vanish
($T_c\rightarrow 0$). On the other hand, a very homogeneous sample has
extended charge carrier states, but $T_c$ is lower since in such case
all the Mn spins are in a similar ``average'' environment.

\subsection{Effects of on-side disorder}

We now consider the role of the on-site disorder $u(i)\ne 0$ in our
model. While the nature of the heavy compensation is not elucidated at
this point, one may assume that compensation is due to the
annihilation of the holes by some type of defect, leading to
appearance of a background of charged impurities. These would create
an electric potential $u(i)$ at all Mn sites. The simplest way to
describe it is to assume that the on-site energies $u(i)$ are
distributed with equal probability in the interval $[-W, W]$. In
Fig. \ref{fig15} we compare the average Mn and spin charge carrier
magnetizations obtained from a random Mn configuration with $x=0.05$
and $p=10\%$, for various values of the on-site disorder cut-off
$W/E_b=0,0.5$ and 1.  On-site disorder does not affect $T_c$
considerably (in all the simulations we performed, we find that $T_c$
decreases slowly with increasing $W$). However, on-site disorder
changes the shape of the magnetization curve. Due to on-site disorder,
some of the charge carriers from the high density regions are pushed
away into the less populated regions. As a result, the magnetization
near $T_c$ (which is dominated by contributions from Mn in the
high-density regions) is suppressed. On the other hand, the
low-temperature magnetization, which is dominated by the spins in the
low density regions, is increased accordingly. It is interesting to
notice that the magnetization of the sample with $W$=1 Ry varies
almost linearly with temperature. Such unusual $M(T)$ dependence has
been observed experimentally. \cite{Esch,Harris}

We have also investigated a more detailed model for on-site disorder,
which assumes that compensation is entirely due to As antisites. When
a valence-V As atom substitute for a valence-III Ga atom, its two
extra electrons effectively remove two holes from the impurity
band. If all the compensation is due to such processes, then the
number of As antisites must be given by $N_{As}=N_{d}(1-p)/2$. Each
such As impurity has an effective charge $+2e$, and therefore will
contribute an on-site Coulomb potential $+2e^2/\epsilon r$ at a Mn
impurity site which is at a distance $r$ from it. However, since the
Mn ions also have effective ionic charge $-e$, the As potential is
screened (partially compensated) by the potential of the Mn impurities
nearby it. One could use a detailed formula
$u(i)=-\sum_{j}^{}e^2/\epsilon r_{j,Mn} + \sum_{k}^{}2e^2/\epsilon
r_{k,As}$, with the first term describing the Mn contribution and the
second one describing the As antisite contribution. An alternative,
simpler form, is to assume that each As antisite only contributes to
the on-site potential $u(i)$ of its two nearest Mn neighbors, with the
contribution to the other Mn sites being screened out by the
contribution of these two nearest Mn sites. The two formulations are
qualitatively equivalent. Given the absence of more detailed
information about the exact nature of compensation and screening, we
investigated the simpler model. In this case, after we randomly choose
the locations for the Mn impurities, we select random positions on the
Ga sublattice for the As antisites as well.  We find the two closest
Mn neighbors for each As antisite (with each Mn selected as neighbor
only for its closest As antisite), compute the corresponding values
for $u(i)$ and then proceed with the calculation as described
previously. Typically, the on-site interaction in this model leads to
a substantial decrease in $T_c$ (see Fig. \ref{fig_rand}). Also, the
shape of magnetization curves becomes even more concave, with the
larger change for the hole magnetization, which no longer reaches full
polarization in the $T=0$ limit.

\subsection{Effects of external magnetic fields}

Finally, we consider the effect of an external magnetic field, in the
absence of on-site interactions $u(i)=0$. We assume that $\tilde{g}=2$
for the Mn spins. The precise value of the g-factor for the holes is
not important, since we find that the magnetization is not changed if
we vary $g$ within a reasonable range. This is a consequence of the
fact that each hole strongly interacts with many Mn spins, and the
external magnetic field is just a small perturbation to the effective
on-site energy $\epsilon_{i \sigma}$ experienced by holes [see
Eq. (\ref{9})]. On the other hand, the external magnetic field leads
to a significant change in the effective magnetic field $H_i$ [see
Eq. (\ref{6})] of each Mn spin, since any Mn interacts with very few
holes (or practically none, for weakly interacting Mn). In
Fig. \ref{fig16} we plot the magnetization of a disordered
configuration with $x=0.05$ and $p=10\%$, in the presence of an
external magnetic field $H=0,5$ and 10T. The external magnetic field
leads to a significant increase of the Mn spin magnetization at all
temperatures, since it polarizes the many weakly interacting spins. It
also leads to a saturation of the magnetization for temperatures $k_BT
< g \mu_BH$, as expected.  The magnetization of the charge carriers is
also increased (in magnitude) in the presence of the magnetic
field. This may seem puzzling at first, since one would expect that
the external magnetic field would favor a flip of the charge carrier
spin from $\sigma=\downarrow$ to $\sigma= \uparrow$, leading to a
decrease of the charge carrier magnetization. However, this is again
due to the fact that each hole interacts with many Mn spins. As the
magnetization $S_{Mn}(i)$ of each Mn spin is increased by the magnetic
field, the effective negative magnetic field felt by the holes
increases, more than compensating the positive external magnetic field
H [see Eq. (\ref{9})]. Addition of random on-site disorder changes the
shape of the magnetization curves, in a similar fashion to the one
presented in Fig.\ref{fig15}.

\subsection{Beyond the hydrogenic model}

As emphasized in the introduction, the impurity band model with
hydrogen-like, exponentially decaying hopping parameters, leaves out
many aspects of the true Hamiltonian in a system like
Ga$_{1-x}$Mn$_x$As, especially with large As antisite defects, or
similar causes for the large observed compensation.  First, and
probably foremost, is that the Mn dopant is not exactly "shallow",
with a binding energy of over 100 meV.  This suggests that the true
wavefunction is substantially affected by central cell corrections and
spin effects, which could affect the hopping integrals
considerably. Secondly, the magnitude of $t(r)$ is based on a
two-center formalism for spherically symmetric wavefunctions valid at
intermediate to large separations. Besides the obvious complication of
anisotropy of the true hole wavefunction, these could lead to
substantial renormalization of the effective $t(r)$ at the densities
of interest, especially at the upper end ($\sim$ few percent).  More
microscopic calculations\cite{BhattRice81} suggest a significant
renormalization of the energies within the impurity band, compared to
the simple two-center tight-binding picture.  Consequently, we discuss
the effects of changing the hopping parameter $t(r)$ in some detail
below.  The effect of other approximations made in our study, namely,
the neglect of the random potential due to the compensating centers,
carrier-carrier interactions, and the valence band states, are
discussed following that.

The effect of reducing the magnitude of the hopping can be achieved by
changing the Bohr radius, or the prefactor. A change in Bohr radius is
merely a renormalization of the effective density, while the effect of
changing the prefactor is tantamount to changing the exchange coupling
in the opposite direction, and then rescaling the temperature scale
appropriately.  To study the effects of restricting the hopping to
nearby sites only, we have studied a model where hopping is limited to
sites within a radius $r_h$. One would expect the parameter $r_h$ to
decrease with density, so as to maintain a reasonable coordination
number, $z= (4\pi/3) n_{Mn} r_h^3$. We use a phenomenological formula
$z= 12n_{Mn}/(n_c+n_{Mn})$, where $n_c$ is the Mn concentration for
which the density of holes $n_{MI}=pn_c$ corresponds to the
metal-insulator transition $n_{MI}^{1/3} a_B=0.25$.  This formula has
the proper asymptotic limits that $z$ is proportional to $n_{Mn}$ at
low density, saturates to $z = 12$ at large $n_{Mn}$ and at the
Metal-Insulator Transition the average coordination number is $z=6$,
the same low coordination number as in the simple cubic lattice, and which
is know to give a reasonable value for the Metal-Insulator Transition
for hydrogen centers.\cite{BhattRice81}  We obtain
$r_h=2.22 a_B$ for $x=0.0092$ corresponding to an average $z=4.7$,
whereas for $x=0.05$ we have $r_h=1.59 a_B$ and $z=9.23$.

Another issue concerns the sign of the hopping integral. In the model
studied, the hopping integrals have been taken to have the same
sign. However, the real system is heavily compensated due perhaps to
As antisite defects, or to some other, as yet unknown, source. In any
case, there will be potentials due to these compensating centers,
which because of their opposite sign, may render some hopping elements
of the other sign. In an effort to see the influence of such effects,
we have also studied models where the hopping integrals have random
sign.

Figure \ref{fighop} shows the magnetization curves obtained for
various models, as described in the figure caption.  As can be
expected, details of the hopping parameter lead to different $T_c$;
consequently we show the data on a scaled plot in terms of $T/T_c$.
The crucial issue appears to be the density of states near the Fermi
level (on the scale of the transition temperature $T_c$). Impurity
bands that are broad have low $T_c$, while those that are relatively
narrow, as might be expected for centers that have a strong central
cell or on-site exchange energy, have high $T_c$.

Despite the variation in $T_c$, we find a number of features that are
generic to the models studied, as illustrated in Fig. \ref{fighop} :

(a) For each model $t(r)$ studied, {\em within the mean field
approximation}, the $T_c$ of the disordered system is higher than that
of the corresponding ordered superlattice, while the magnetization at
low temperatures ( below $T_c$/3 or so ) is lower for the disordered
system.

(b) The magnetization curves obtained with impurity bands from a set
of positional disordered impurity centers have an unusual shape
(concave upwards), or at least significantly less concave downwards
than the standard concave downward (convex upward) form of most
uniform magnets.

(c) There is significant temperature dependence of $M(T)$ at
temperatures which are well below $T_c$, unlike in essentially uniform
magnetic models where the magnetization has reached its $T=0$
saturation value at $T_c/2$, or certainly by $T_c/3$.

(d) Some of the unusual features of the Mn spin curves, which
determine the bulk magnetization, are seen in the hole magnetization
curves as well (see Fig.\ref{fighop}) , but they are less pronounced.

Other approximations made in our study involve neglecting a number of
complications present in the experimental system such as: (i) the
effect of the compensating centers, (ii) carrier-carrier interactions,
and (iii) the valence band states. For (i), we have considered
(subsection D) one of the effects, namely, if we include the random
potential from these charged centers as an effective on-site random
potential in our tight-binding Hamiltonian for the extreme cases (a)
where the onsite potential is uncorrelated, and (b) where it is
modeled as being due to a close by As antisite defect.  However, the
longer range nature of the Coulomb potential in these systems with
poor screening could also affect the hopping parameters, as we
discussed in the preceding paragraphs.

The states near the Fermi level for the range of concentrations we
consider are not strongly localized in the model studied, and even the
actual system is fairly close to a metal-insulator transition (which
we view as primarily driven by disorder, not electron correlation).
Consequently, we believe that approximation (ii) is reasonable. An
on-site interaction $U$ (as in a Hubbard model), if treated in a
mean-field approximation, would lead to increased $T_c$, since it aids
in the splitting of the up and down spin bands. We show results for
one such study that we carried out in Fig \ref{figU} which confirms
the above expectation. The change in $T_c$ is about 35 \% for the
$x=0.01$ case and 18\% for the $x=0.05$ case, for a value $U$ = 1 Ry
typical for hydrogenic centers at low densities. However, this is
likely an overestimate, since a fair fraction of the Mn impurity
binding energy comes from short range potentials and exchange, and
further, the effective $U$ near the metal-insulator transition is
likely to be reduced by screening processes.

Finally, we discuss the neglect of the valence band states. In the
model we studied, we believe it is not important, at least at low
temperatures, where the shape of the magnetization curves is anomalous
and where the mean field results should be at least qualitatively
correct. This is because the Fermi level lies $\sim$ 300 meV above the
valence band minimum, and so excitation to the valence band states
would become dominant only at higher temperatures. It should be
emphasized that the impurity states that we consider are derived from
the host band (this would be the valence band in the case of
Ga$_{1-x}$Mn$_x$As). Consequently, if one wishes to include the rest
of the band states of the host, they must be orthogonalized to the
impurity states; this has the effect of pushing the host band further
from the impurity band, making the effect of host band states smaller
than might be normally expected. In a more realistic model, however,
this remains an open question.

\section{Summary and Conclusions}

In this study we analyzed a simplified model of III-V Diluted Magnetic
Semiconductors, in which the charge carriers are restricted to a band
formed from impurity orbitals, as in shallow doped semiconductors.  We
believe that this is a good starting point, because the high
compensation present in these systems leads to carrier densities which
are not large enough to screen out the Coulomb interaction between the
Mn ions and the charge carriers. This is clearly indicated by the
presence of Metal-Insulator Transitions within the range of doping
densities studied ($x$= 1-5\% Mn).

We started by analyzing an ordered superlattice case, with a
homogeneous charge carrier distribution. This allows us to understand
the rather unusual shape of the magnetization curves found especially
for low values of $p$.  We find good agreement with experimentally
measured $T_c$ with nominal parameters given in the
literature. However, we believe this is somewhat fortuitous, as it is
well known in most magnetic models that mean-field analyses
significantly overestimate the true $T_c$. Given the rather large
number of free parameters, the experimental uncertainties about their
exact values and the exponential dependencies on some of them in our
model, it would always be possible to obtain good agreement with the
experimentally measured $T_c$ by changing the input parameters within
their error bars.

We then show that positional disorder and on-site random interactions
lead to significant changes in the shape of the magnetization and the
critical temperature $T_c$, at least within a mean-field
approximation. This enhancement of the critical temperature by
disorder has been confirmed, with different techniques, in two recent
studies of the III-V DMS.\cite{Mill,Chud} The critical temperatures
obtained within our mean-field scheme for the same input parameters
are larger than the ones measured experimentally by about $50\%$ for
the $x=0.05$ sample, and about a factor of 3 for the $x=0.01$ sample.
While this may appear problematic, it is well known that mean-field
schemes often significantly overestimate the true $T_c$, as we discuss
later. It should be emphasized that the exchange coupling we have
used, obtained from Bhattacharjee et al. \cite{BG} is lower than the
values used by MacDonald et al. \cite{MacD} or Millis and Das Sarma
\cite{Millis} for obtaining a similar $T_c$. The reason for this is
that the impurity wavefunctions that give rise to our impurity band
are peaked at the Mn sites, and therefore provide a greater charge
density at the Mn site than do host band wavefunctions.

We have not performed detailed numerical fits to $T_c$ because of a
number of factors, such as (a) the complicated nature of the hole
wavefunctions, (b) uncertainties about the nature of the compensation
processes, (c) uncertainties about the precise description of the
hopping integral, (d) the under-estimation of the effect of thermal
fluctuations by the mean-field approximation, etc.  One of the robust
results that follows from our study is that the magnetization curves
$M(T)$ may vary considerably from the canonical concave downward
(convex upward) form seen in practically all uniform magnetic models,
independent of dimension or spin-components.  Using different hopping
parameters, we find that while $T_c$ changes with the model chosen,
the concave-upward form of $M(T)$ over much of the temperature range
below $T_c$ is dependent mostly on the Mn spin concentration $x$ and
the carrier density $px$. The curves vary from essentially concave
(convex upwards) functions (as reported in Ref. 1) for large relative
charge carrier concentrations $p$, to almost linear dependence (as
reported in Ref. 4), to very concave upward functions (as reported in
Ref. 10) as $p$ is decreased.  Below $T_c$, our calculated
magnetization curves generically show a fast increase, followed by
saturation and then fast increase again at much lower
temperature,similar to the one reported in Ref. 19.  Overall, we claim
to find good qualitative agreement with the experimental behavior
concerning possible shapes of magnetization curves, the
metal-insulator transition etc. Our study suggests that by appropriate
tuning of various parameters, one may tailor the magnetic behavior
M(H,T) in a manner not possible in simple uniform
magnets.\cite{Kaneyoshi} We believe that detailed information provided
by experiments (using local probes such as ESR and NMR) would allow a
clearer understanding of the nature of ferromagnetism in these
compounds.

Returning to the issue of the magnitude of $T_c$, we believe the most
important correction to our mean-field result is due to thermal
(temporal) fluctuations, which are not included in a mean-field
treatment. This can cause a substantial decrease in the critical
temperature $T_c$. While typical renomalizations of $T_c$ for uniform
models range from tens of percent to factors of 2 or so, we expect
them to be significantly larger for models with great spatial
inhomogeneity, where percolation aspects may have considerable
influence. Preliminary Monte Carlo simulations\cite{malcolm2} which
include these fluctuations show that the the critical temperature is
significantly suppressed with respect to the mean-field value,
especially for the lower concentrations $x$, where we found that the
magnetization is somewhat less robust.

In addition to fluctuation effects, there are several interactions not
included in our model which may lead to an overall decrease of $T_c$
{\em even within mean-field}.  One of these is the hole-hole Coulomb
repulsion, which could prevent the accumulation of all the charge
carriers in the high density regions, and would favor a more uniform
spreading of the charge carriers over the entire sample.  This would
result in a smaller enhancement of $T_c$ in the positionally
disordered model vis-a-vis an ordered superlattice of magnetic ions,
and thus lead to a lowering of $T_c$. However, the magnitude of this
effect may be small because the density of charge carriers is low.

Another possible interaction not included in our simple model, that
may be more effective in preventing the concentration of charge
carriers in the high-density regions, is direct Mn-Mn interactions.
Mn-Mn interactions are expected to be AFM, as they are in II-VI
DMS,\cite{Furdyna} while the charge carriers promote effective
ferromagnetic Mn-Mn interactions. As a result, frustration is expected
to appear, and to be most significant in the high-density regions
where Mn-Mn interactions would be largest. In particular,
nearest-neighbor Mn impurities may lock in a singlet state, and
therefore not contribute at all to magnetization. To first
approximation, one may equate a system with such singlets to a system
whose effective Mn concentration is smaller than the nominal one (the
difference being the concentration of singlets), and which is
restricted to not having any nearest-neighbor Mn sites. In other
words, it is as if these singlets become invisible, as far as the
magnetic properties of the system are concerned.  Another phenomenon
that may be responsible for an effective homogenization of the sample
is the creation of MnAs clusters, which is thought to be the reason
behind the saturation of $T_c$ for $x > 0.05$.\cite{Esch} Although one
can dope more Mn into the sample, an increased concentration $x$ does
not necessarily imply an increased number of magnetically and
electrically active centers. In fact, some of the Mn impurities form
small disordered six-fold coordinated centers with As, as observed in
the closely related InMnAs compound. \cite{Mune} Such centers are
believed to give rise to $n$-type conductivity; in other words, they
provide one possible compensation mechanism. These Mn impurities do
not participate in the AFM exchange with the charge carriers,
described above. One expects that the probability of appearance of
such complexes is higher in the high-density Mn regions.  Appearance
of such complexes in the high-density regimes would lead to an
effective decrease of the local active Mn density, and would
effectively promote again a more homogeneous ground-state, leading to
a lower $T_c$.  Finally, as the temperature increases towards $T_c$,
one expects that charge carriers may be excited to the valence band
states (whose existence has been neglected in this model). This is
likely to lead to a decrease of $T_c$, because the Bloch states would
(a) lead to a more uniform distribution of carriers, and (b) have less
amplitude at the Mn sites than the impurity states. The magnitude (and
importance) of this effect would, however, depend on the density of
states of the impurity band and its separation from the host band.

In conclusion, the nature of ferromagnetism in doped DMS is strongly
affected by disorder. Surprisingly, we find that disorder actually
results in a higher $T_c$. Whether this is due to the mean-field
approximation, or is a more robust phenomenon, awaits results of Monte
Carlo simulation studies\cite{malcolm2}.  Nevertheless, the
versatility provided by a magnetization curve (and ensuing
thermodynamic properties) that has a tunable shape, makes DMS
ferromagnetism a very interesting problem from a theoretical point of
view.  Adding to the richness are possible effects of direct Mn-Mn
interactions in concentrated systems (which lead to spin-glass
behavior in undoped II-VI DMS\cite{Furdyna}), the existence of a
ferromagnetic metal-insulator transition (unlike conventional doped
semiconductors and amorphous alloys), and the likely unusual electron
and spin transport characteristics because of disorder.

\section*{Acknowledgments}
We acknowledge many useful discussions with Malcolm P. Kennett.  This
research was supported by NSF DMR-9809483.  M.B. acknowledges support
from a Natural Sciences and Engineering Research Council of Canada
Postdoctoral Fellowship.  R. N. B. acknowledges the hospitality of the
Aspen Center for Physics, while this manuscript was in preparation.

\newpage

\begin{figure}
\centering
\parbox{0.45\textwidth}{\psfig{figure=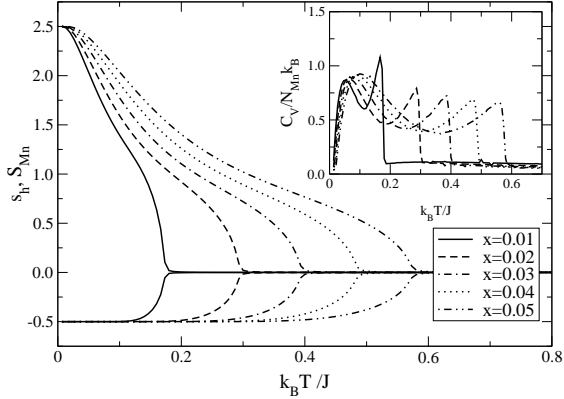,width=85mm,angle=270}}
\caption{\label{fig1} The average Mn spin $S_{Mn}$ and the average
spin per charge carrier $s_h= \sum_{i}^{} s_h(i) / N_h = s_{hole}/p $
for doping concentrations $x=0.01,0.02,0.03,0.04$ and 0.05, and
$p=10\%$.  Due to their antiferromagnetic interaction, the two
expectation values have opposite sign, with the Mn spin saturating at
${5\over 2}$ and the charge carrier spin saturating at $-{1\over 2}$
at low temperatures. The Mn spin curves have an uncharacteristic
shape, with inflection points as explained in the text.  In the inset
we plot the specific heat per Mn impurity as a function of
temperature. The lower peak is due to Mn spin fluctuations, while the
upper peak is due to charge carrier fluctuations.  }
\end{figure}

\begin{figure}
\centering
\parbox{0.45\textwidth}{\psfig{figure=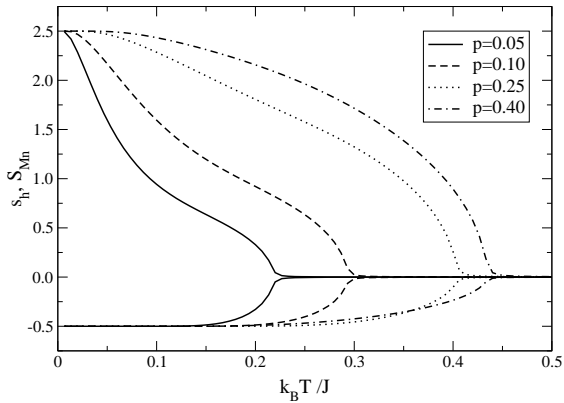,width=85mm,angle=270}}
\caption{\label{fig2} Evolution of the average Mn spin $S_{Mn}$ and
the average spin per charge carrier $s_h$ for different charge carrier
concentrations $p=5,10,25$ and $40\%$. The Mn concentration is fixed
at $x=0.02$. With large $p$, the shape of $S_{Mn}$ becomes more like
the usual convex upward function seen in typical
ferromagnets. However, for low doping concentrations $p$ the
relatively few charge carriers can only fully magnetize all the Mn
spins at very low temperatures.  }
\end{figure}

\begin{figure}
\centering
\parbox{0.45\textwidth}{\psfig{figure=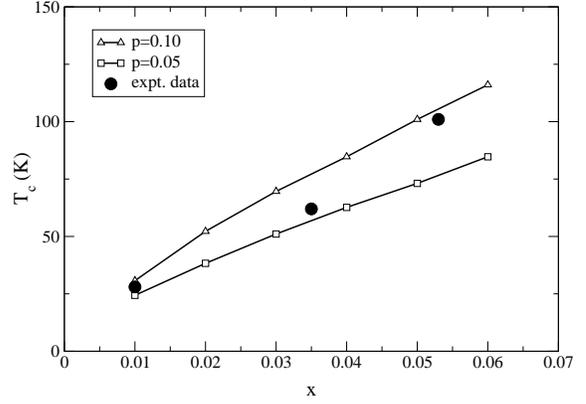,width=85mm,angle=270}}
\vspace{5mm}
\caption{\label{fig3} Comparison between experimentally measured
critical temperatures $T_c$ (full circles, from Ref. 10), and values
obtained in the mean-field approximation for the ordered Mn spin
case. The experimental points fall between the lines corresponding to
$p=5\%$ and $p=10\%$, which is the estimated range of charge carrier
concentration for these crystals.\cite{Besch} However, the $T_c$
values are very strongly dependent on $p$ (see Fig.\ref{fig2}) as well
as other parameters of the model. }
\end{figure}

\begin{figure}
\centering
\parbox{0.45\textwidth}{\psfig{figure=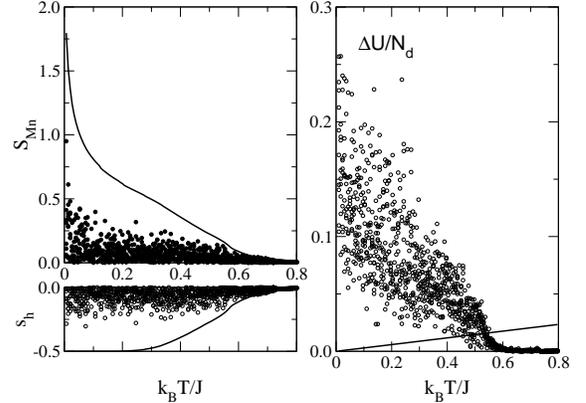,width=85mm,angle=270}}
\vspace{5mm}
\caption{\label{fig5} Left panel: Comparison between values for the
average Mn spin $S_{Mn}$ and the average charge carrier spin $s_h$
obtained with biased initial conditions (full line) and random initial
conditions (circles). For random initial conditions we actually plot
$|S_{Mn}| > 0$ (full circles) and $-|s_h|<0 $ (empty circles), since
the two expectation values always have opposite sign, but the
orientation is arbitrary. Results are for $x=0.00926$, $p=10\%$. Right
panel: the difference in energy per Mn spin between random and biased
initial configurations, as a function of temperature. Full line is
$k_BT/N_d$. For details, see text.}
\end{figure}

\begin{figure}
\centering
\parbox{0.45\textwidth}{\psfig{figure=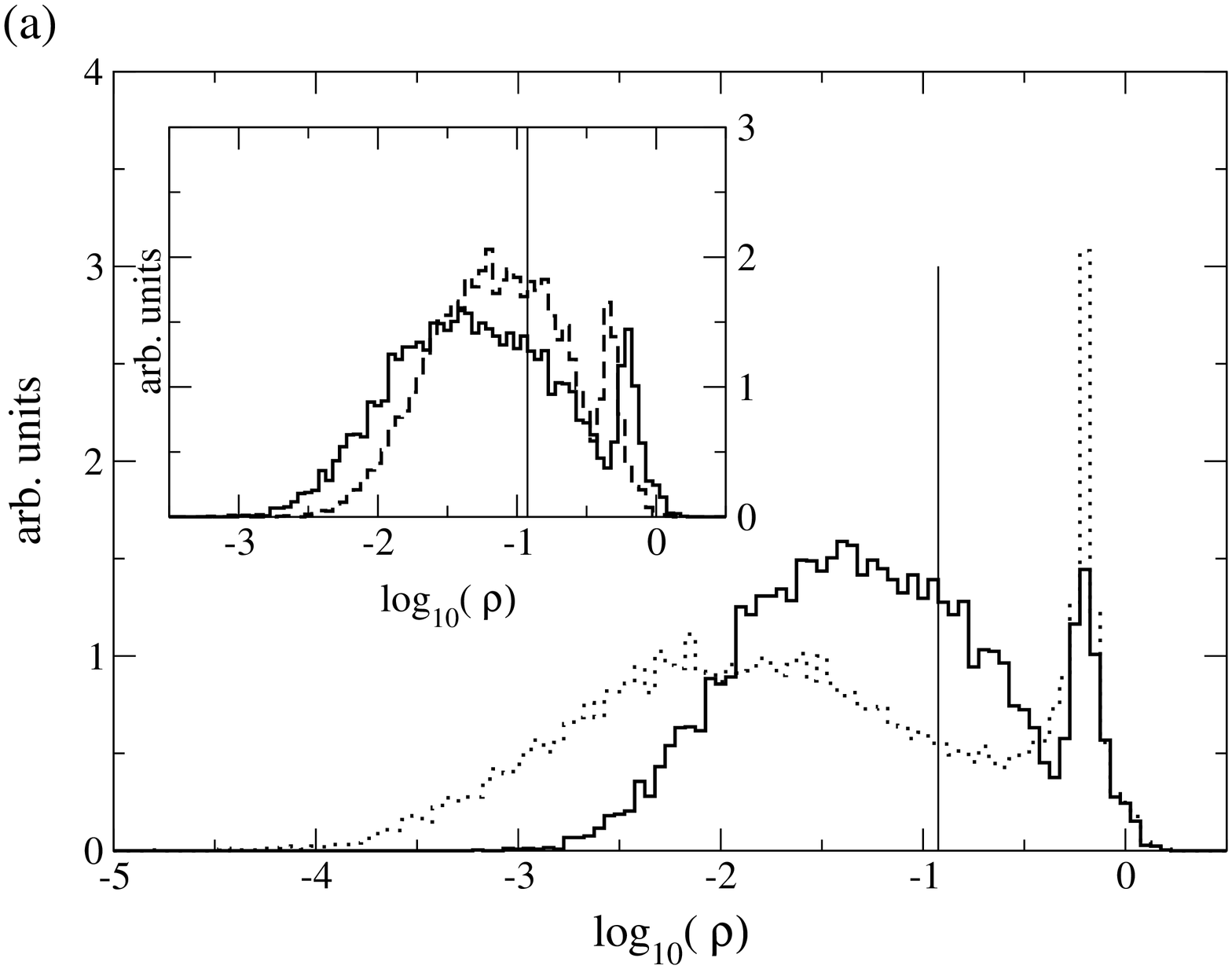,width=85mm,angle=270}}
\parbox{0.45\textwidth}{\psfig{figure=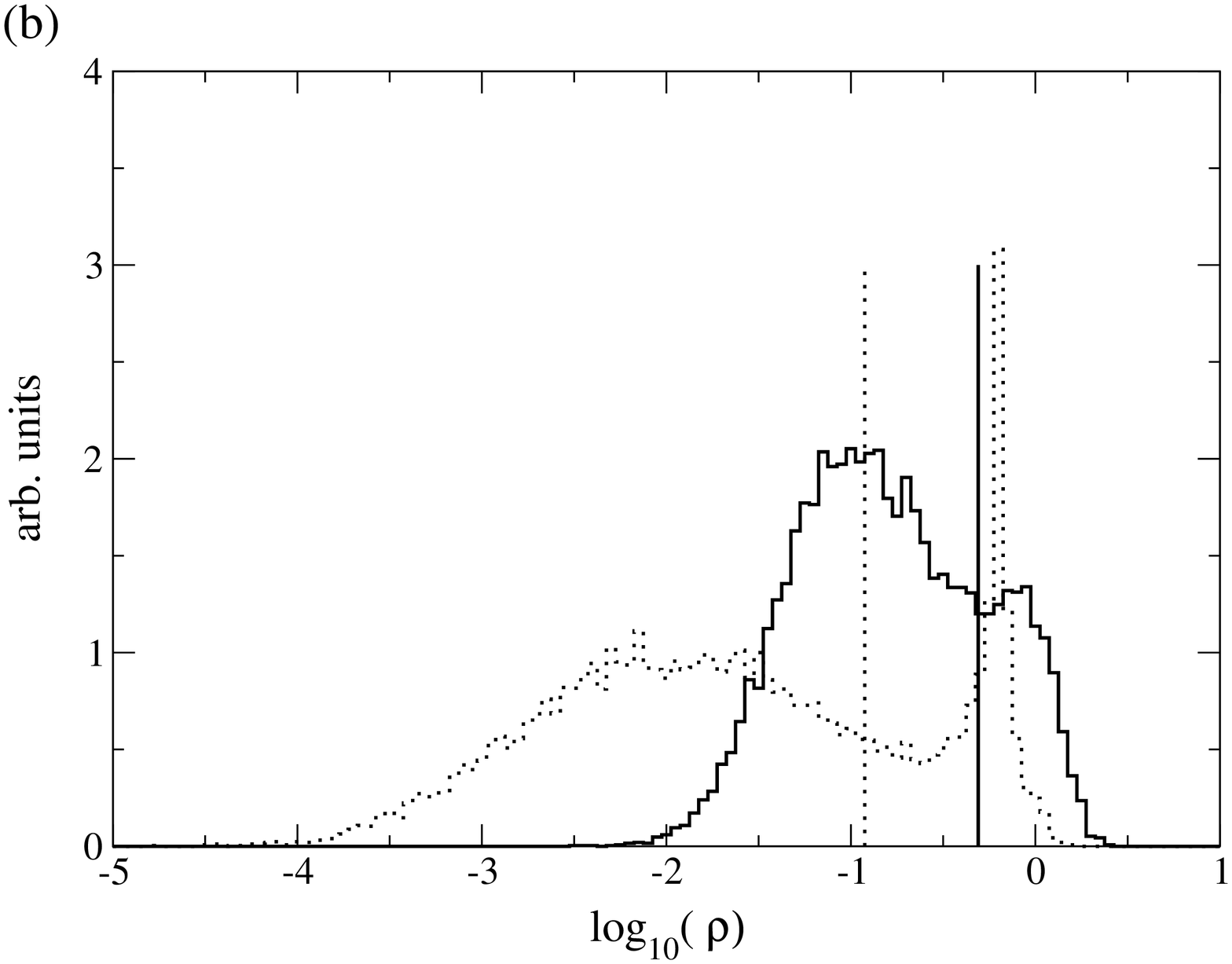,width=85mm,angle=270}}
\vspace{5mm}
\caption{\label{fig6}(a) Histogram (in arbitrary units) of the total
density of charge carriers at each Mn site $\log_{10}(\rho(i))$, for
random Mn configuration with $x=0.00926$ and $p=10\%$. The dotted line
corresponds to $k_BT/J=0.0006$, while the full line corresponds to
$k_BT/J=0.6$.  The vertical line indicates the position of the
$\delta$-function that describes the same histogram for an ordered Mn
lattice. The two-peaked structure of the histogram for the disordered
samples shows that in this case some Mn interact with holes much more
strongly than the average, while some Mn spins interact with holes
much more weakly than the average. Inset: Same, but for temperatures
$k_BT/J=0.6$ (full line) and $k_BT/J=1.5$ (dotted line).  (b)
Comparison between the density of charge carrier histogram for
$x=0.00926$ (dotted line) and $x=0.05$ (full line) systems. Both
systems have $p=10\%$ and $k_BT/J=0.0005$.  Vertical lines show the
corresponding values for the ordered systems. The double-peaked
structure is also visible in the higher density sample.  }
\end{figure}

\begin{figure}
\centering
\parbox{0.45\textwidth}{\psfig{figure=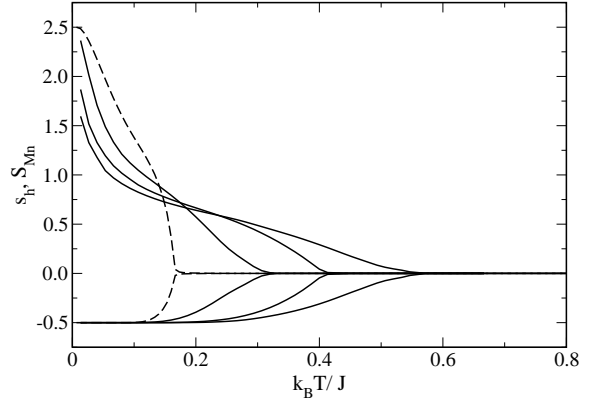,width=85mm,angle=270}}
\vspace{5mm}
\caption{\label{fig7} The average Mn spin $S_{Mn}$ and average spin
per hole $s_{h}$ for doping concentration $x=0.00926$ and $p=10\%$. In
increasing order of $T_c$, the curves correspond to ordered, weakly
disordered, moderately disordered and completely random distributions
of Mn (see text). }
\end{figure}

\begin{figure}
\centering
\parbox{0.45\textwidth}{\psfig{figure=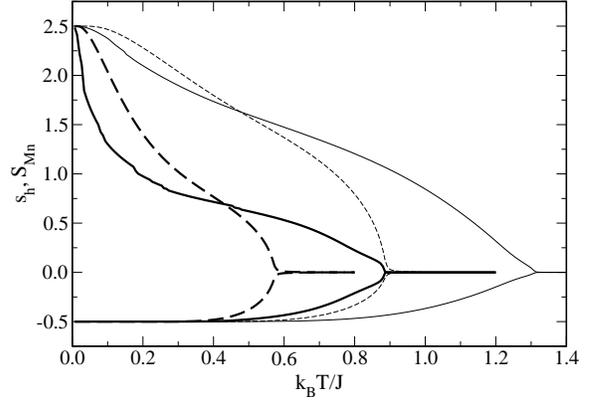,width=85mm,angle=270}}
\vspace{5mm}
\caption{\label{fig8}The average Mn spin $S_{Mn}$ and the average spin
per hole $s_{h}$ for doping concentration $x=0.05$ and $p=10\%$ (thick
lines) and $30\%$ (thin lines) for typical random Mn distributions
(full lines) and simple cubic ordered Mn distributions (dashed lines).
}
\end{figure}

\begin{figure}
\centering
\parbox{0.45\textwidth}{\psfig{figure=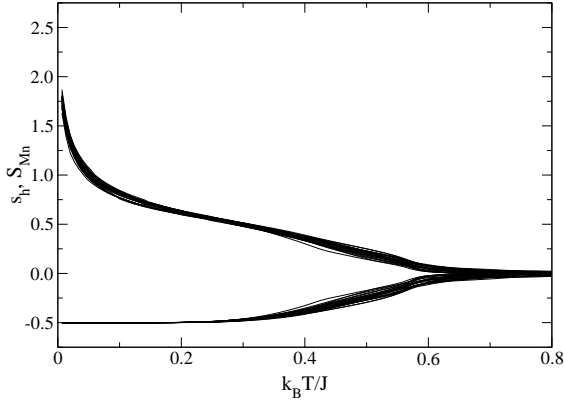,width=85mm,angle=270}}
\parbox{0.45\textwidth}{\psfig{figure=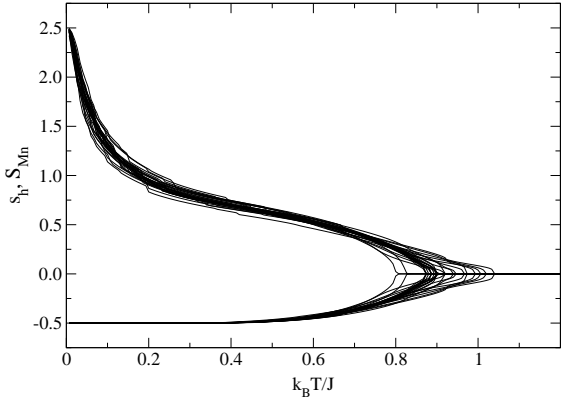,width=85mm,angle=270}}
\vspace{5mm}
\caption{\label{fig9} Magnetizations for 25 realizations of Mn
positional disorder, for $x=0.00926$ (upper panel) and $x=0.05$ (lower
panel). The long tails near $T_c$ (especially for the lower density)
are due to polarization of a few very dense clusters of Mn, and would
be destroyed by thermal fluctuations.}
\end{figure}

\newpage

\begin{figure}
\centering
\parbox{0.45\textwidth}{\psfig{figure=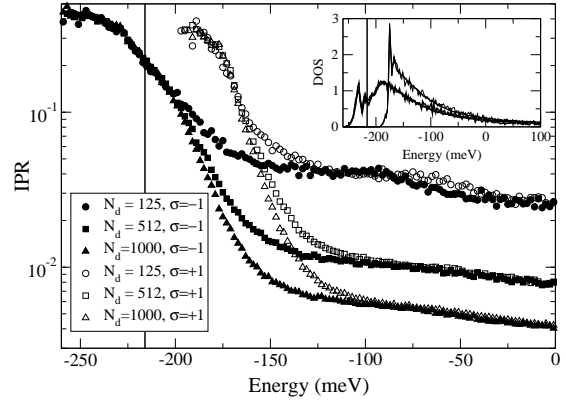,width=85mm,angle=270}}
\vspace{5mm}
\caption{\label{fig10} Average Inverse Participation Ratio as a
function of energy for the $\sigma=\downarrow$ (full symbols) and
$\sigma=\uparrow$ (empty symbols) sub-bands of completely random Mn
distributions with $p=10\%$, $x=0.00926$ and $N_d=125$ (circles), 512
(squares) and 1000 (triangles). The vertical line shows the Fermi
energy. Clearly, all the occupied states are localized. In the inset
we plot the total Density of States for the two subbands, in arbitrary
units.  }
\end{figure}

\begin{figure}
\centering
\parbox{0.45\textwidth}{\psfig{figure=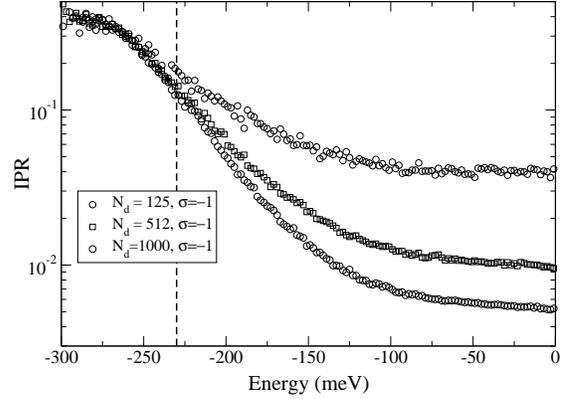,width=85mm,angle=270}}
\vspace{5mm}
\caption{\label{fig11} Average Inverse Participation Ratio as a
function of energy for the $\sigma=\downarrow$ subband of systems with
strong disorder (full symbols) and medium-disorder (empty symbols),
for $p=10\%$, $x=0.00926$ and $N_d=125$ (circles), 512 (squares) and
1000 (triangles). The vertical line shows the Fermi energy of the
system with strong disorder (full line) and medium disorder (dashed
line). For these parameters, a medium-disordered sample is just above
the MIT, while a strongly disordered one is just below the MIT.  }
\end{figure}

\begin{figure}
\centering
\parbox{0.45\textwidth}{\psfig{figure=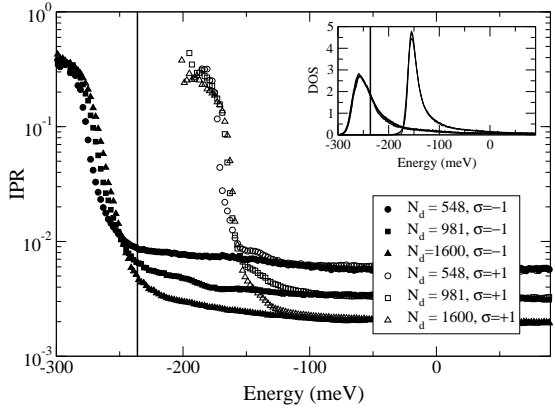,width=85mm,angle=270}}
\vspace{5mm}
\caption{\label{fig12} Average Inverse Participation Ratio as a
function of energy for the $\sigma=\downarrow$ (full symbols) and
$\sigma=\uparrow$ (empty symbols) subbands of completely random Mn
distributions with $p=10\%$, $x=0.05$ and $N_d=548$ (circles), 981
(squares) and 1600 (triangles). The vertical line shows the Fermi
energy, above the mobility edge. In the inset we plot the total
Density of States for the two subbands, in arbitrary units.  }
\end{figure}

\begin{figure}
\centering
\parbox{0.45\textwidth}{\psfig{figure=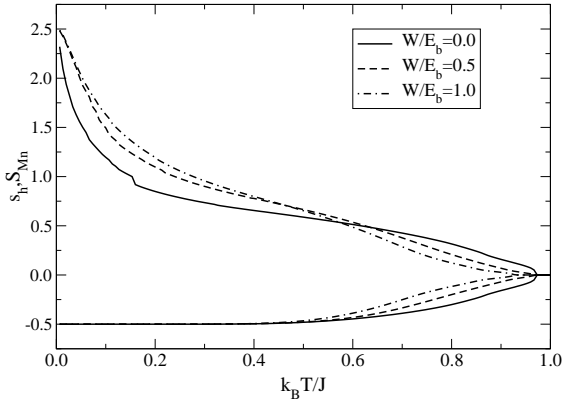,width=85mm,angle=270}}
\vspace{5mm}
\caption{\label{fig15} Mn spin and charge carrier average
magnetization as a function of temperature, for one random Mn
configuration corresponding to $x=0.05$ and $p=10\%$, and different
values of the on-site disorder cut-off $W/E_b=0,0.5$ and 1.While
on-site disorder does not affect $T_c$ considerably, it does change
the shape of the magnetization curves.}
\end{figure}

\begin{figure}
\centering
\parbox{0.45\textwidth}{\psfig{figure=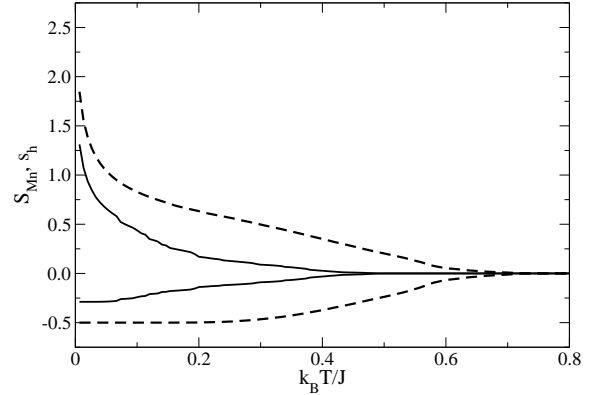,width=85mm,angle=270}}
\vspace{5mm}
\caption{\label{fig_rand} Comparison between magnetization curves in
the absence of an on-site interaction $u(i)=0$ (dashed line), and with
an on-site interaction due to As antisites, as explained in text (full
line). The curves correspond to the same disordered positions for the
Mn sites, and $x=0.00924$ and $p=10\%$.}
\end{figure}

\begin{figure}
\centering
\parbox{0.45\textwidth}{\psfig{figure=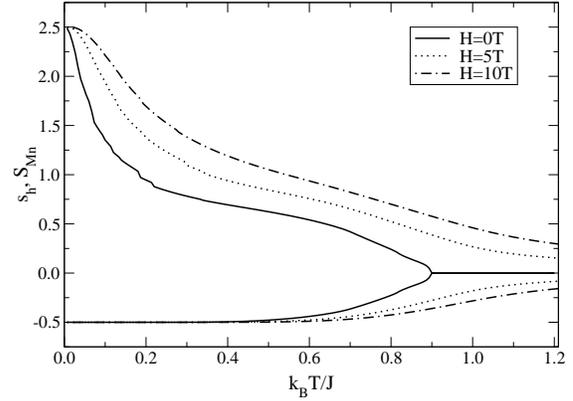,width=85mm,angle=270}}
\vspace{5mm}
\caption{\label{fig16} Mn spin and charge carrier average
magnetization as a function of temperature, for one random Mn
configuration corresponding to $x=0.05$ and $p=10\%$, for an external
magnetic field $H=0,5$ and 10T. The overall magnetization is
significantly increased at lower temperatures, since the external
magnetic field polarizes all the Mn spins, not only the ones in the
strongly interacting regions.}
\end{figure}

\begin{figure}
\centering
\parbox{0.45\textwidth}{\psfig{figure=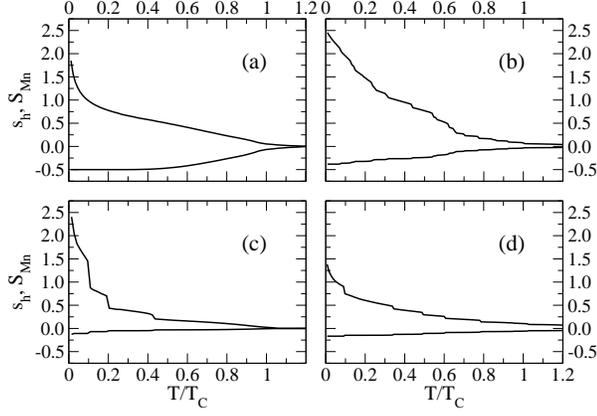,width=85mm,angle=270}}
\caption{\label{fighop} Mn spin and charge carrier average
magnetization as a function of $T/T_c$, for {\em one} random Mn
configuration corresponding to $x=0.00926$ and $p=10\%$. The curves
correspond to: (a) hopping allowed between all sites, same-sign
hopping integral ($T_c/J\sim 0.6$); (b) hopping allowed only for sites
with $r_h=2.22a_B$ of each other, same-sign hopping integral
($T_c/J\sim 0.2$); (c) hopping allowed between all sites, random-sign
hopping integral ($T_c/J\sim 0.12$); (d) hopping allowed only for
sites with $r_h=2.22a_B$ of each other, random-sign hopping integral
($T_c/J\sim 0.2$). Sample averaging is required to obtain smooth
curves.  For more details, see text. }
\end{figure}

\begin{figure}
\centering
\parbox{0.45\textwidth}{\psfig{figure=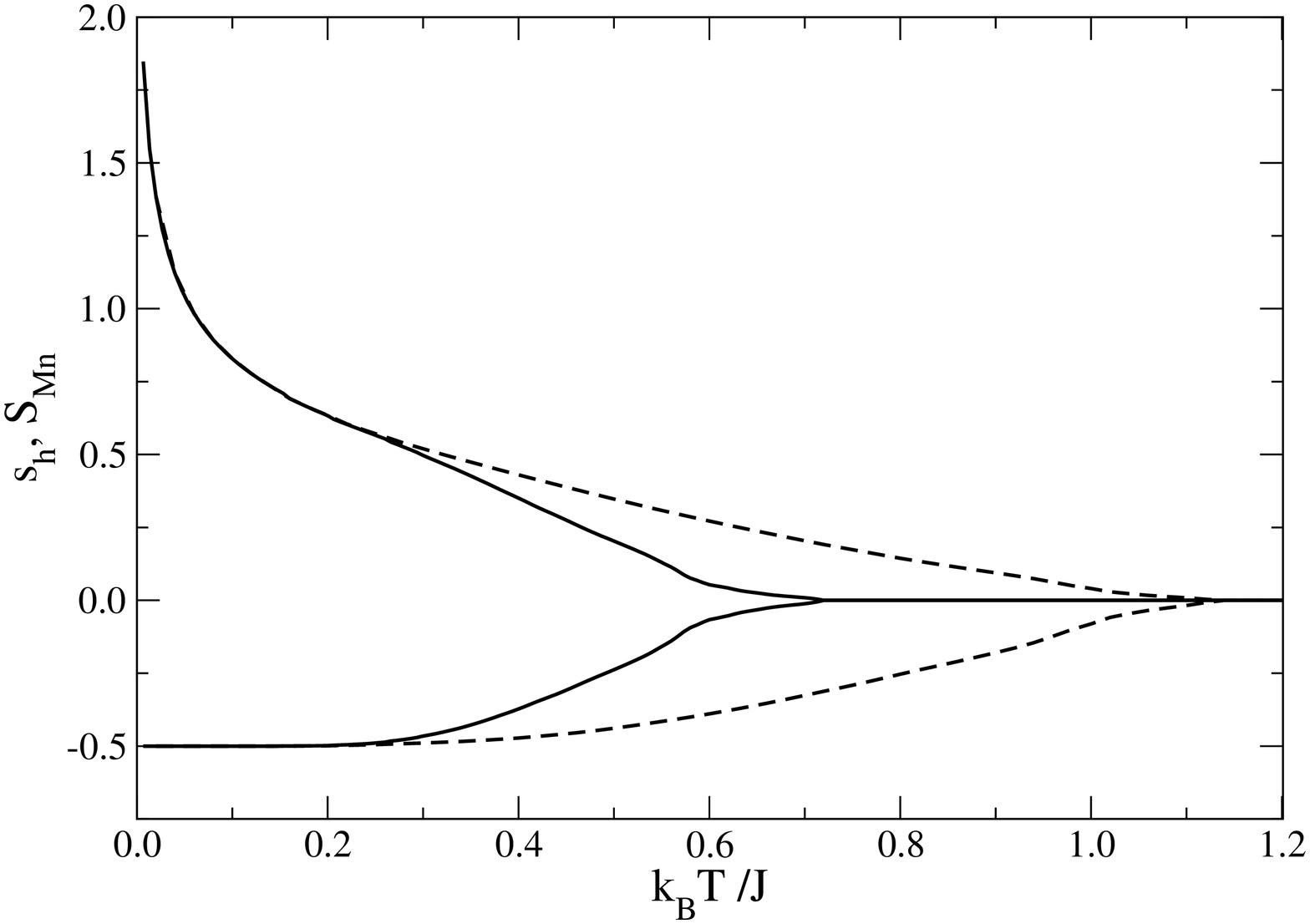,width=85mm,angle=270}}
\caption{\label{figU} Effect of an on-site interaction $U$ = 1 Ry
(treated within the mean-field approximation) on the magnetization
curves corresponding to $x=0.01$ and $p=10\%$ (dashed curve). For
comparison, the curves obtained in the absence of an on-site
interaction, for the same realization of disorder, are shown as well
(full line).  }
\end{figure}


\begin{thebibliography}{99}

\bibitem{Ohnorev} for a review of properties of ferromagnetic III-V
semiconductors, see H. Ohno, J. Magn. Magn. Mat. {\bf 200}, 110
(1999).

\bibitem{Ohno} H. Ohno, A. Shen, F. Matsukura, A. Oiwa, A. Endo,
S. Katsumoto and Y. Iye, Appl. Phys. Lett. {\bf 69}, 363 (1996).

\bibitem{Haya} T. Hayashi, M. Tanaka, T. Nishinaga, H. Shimada,
H. Tsuchiya, Y. Otsuka, J. Cryst. Growth {\bf 175}, 1063 (1997).

\bibitem{Esch} A. Van Esch, L. Van Bockstal, J. De Boeck, G. Verbanck,
A. S. van Steenbergen, P. J. Wellmann, G. Grietens, R. Bogaerts,
F. Herlach and G. Borghs, Phys. Rev. B {\bf 56}, 13103 (1997).


\bibitem{RKKY} T. Dietl, A. Haury and Y. M. d'Aubign\'{e},
Phys. Rev. B {\bf 55}, R3347 (1997); M. Takahashi, Phys. Rev. B {\bf
56}, 7389 (1997); T. Jungwirth, W. A. Atkinson, B. H. Lee and
A. H. MacDonald, Phys. Rev. B {\bf 59}, 9818 (1999); T. Dietl,
H. Ohno, F. Matsukura, J. Cibert and D. Ferrand, Science {\bf 287},
1019 (2000).

\bibitem{MacD} J. K\"{o}nig, H.-H. Lin and A. H. MacDonald,
Phys. Rev. Lett. {\bf 84}, 5628 (2000).

\bibitem{BG} A. K. Bhattacharjee and C. B. \'{a} la Guillaume, Solid
State Comm. {\bf 113}, 17 (2000).

\bibitem{SE} B. I. Shklovskii and A. L. Efros, {\em Electronic
Properties of Doped Semiconductors}, (Springer-Verlag, Berlin, 1984).

\bibitem{BhattRice81} R. N. Bhatt and M. T. Rice, Phys. Rev. B {\bf
23}, 1920 (1981).

\bibitem{Besch} B. Beschoten, P.A. Crowell, I. Malajovich,
D. D. Awschalom, F. Matsukura, A. Shen and H. Ohno, Phys. Rev.
Lett. {\bf 83}, 3073 (1999).

\bibitem{Katsu} S. Katsumoto, A. Oiwa, Y. Iye, H. Ohno, F. Matsukura,
A. Shen, Y. Sugawara, phys. stat. sol. (b) {\bf 205}, 115 (1998).

\bibitem{MB1} Mona Berciu and R. N. Bhatt, Phys. Rev. Lett. {\bf 87},
107203 (2001).

\bibitem{MacDonald} J. Schliemann, J. Konig and A.H. MacDonald,
Phys. Rev. B {\bf 64}, 165201 (2001).
 

\bibitem{Bhatt1} R. N. Bhatt, Phys. Rev. B {\bf 24}, 3630 (1981).

\bibitem{MacDSch} J. Schliemann and A. H. MacDonald, cond-mat/0107573.


\bibitem{Malcolm} Malcolm P. Kennett, Mona Berciu and R. N. Bhatt,
cond-mat/0102315.


\bibitem{oiwa} A. Oiwa, S. Katsumoto, A. Endo, M. Hirasawa, Y. Iye,
H. Ohno, F. Matsukura, A. Shen and Y. Sugawara, Solid State Comm. {\bf
103}, 209 (1997).

\bibitem{Matsu} F. Matsukura, H. Ohno, A. Shen and Y. Sugawara,
Phys. Rev. B {\bf 57}, R2037 (1998).

\bibitem{Harris} J. G. E. Harris, D. D. Awschalom, F. Matsukura,
H. Ohno, K.D . Maranowski and A. C. Gossard, Appl. Phys. Lett. {\bf
75}, 1140 (1999).



\bibitem{Mill} A. J.  Millis, invited talk at the International
Conference on Novel Aspects of Spin-Polarized Transport and Spin
Dynamics, Washington DC, Aug. 9-11, 2001.

\bibitem{Chud} A. L. Chudnovskiy, cond-mat/0108396.

\bibitem{Millis} A. Chattopadhyay, S. Das Sarma and A. J. Millis,
cond-mat/0106455.

\bibitem{Kaneyoshi} Uniform magnets as well as metallic alloys are
characterized by weak deviations from a universal curve on a
$M(H,T)/M(0,0)$ versus $T/T_c$ and $g\mu H/k_BT_c$ plot. [See
T. Kaneyoshi, {\em Introduction to Amorphous Magnets}, p 56-61 (World
Scientific, 1992)].


\bibitem{malcolm2} Malcolm P. Kennett, Mona Berciu and R. N. Bhatt,
unpublished; R. N. Bhatt, X. Wan, M. P. Kennett and M. Berciu,
submitted to Comp. Phys. Comm.

\bibitem{Furdyna} See {\em e.g.} S. Oseroff and P. H. Keesom in {\em
Diluted Magnetic Semiconductors} , edited by J. K. Furdyna and
J. Kossut (Academic Press, 1988).


\bibitem{Mune} H. Munekata, H. Ohno, S. von Molnar, A. Segm\"{u}ller,
L. L. Chang and L. Esaki, Phys. Rev. Lett. {\bf 63}, 1849 (1989).



\end{thebibliography}
\end{document}